\documentclass[twocolumn,showpacs,
prb
,superscriptaddress
]{revtex4-1}

\usepackage{graphicx}
\usepackage{amssymb}
\usepackage{amsmath}
\usepackage[usenames,dvipsnames]{color}
\usepackage{comment}

\newcommand{\vect}[1]{{\mathbf #1}}
\newcommand{\Frac}[2]{\displaystyle\frac{#1}{#2}}

\begin{document}


\title{Probing the collective excitations of a spinor polariton fluid}

\author{M. Van Regemortel}
\email{Mathias.VanRegemortel@uantwerpen.be}
\affiliation{TQC, Universiteit Antwerpen, Universiteitsplein 1, B-2610
  Antwerpen, Belgium}

\author{M. Wouters}
\affiliation{TQC, Universiteit Antwerpen, Universiteitsplein 1, B-2610
  Antwerpen, Belgium} 

\author{F. M. Marchetti} 
\affiliation{Departamento de F\'isica Te\'orica de la Materia
  Condensada \& Condensed Matter Physics Center (IFIMAC), Universidad
  Aut\'onoma de Madrid, Madrid 28049, Spain}

\date{\today}

\begin{abstract}
  We propose a pump-probe set-up to analyse the properties of the
  collective excitation spectrum of a spinor polariton fluid. By using
  a linear response approximation scheme, we carry on a complete
  classification of all excitation spectra, as well as their intrinsic
  degree of polarisation, in terms of two experimentally tunable
  parameters only, the mean-field polarisation angle and a rescaled
  pump detuning. We evaluate the system response to the external
  probe, and show that the transmitted light can undergo a spin
  rotation along the dispersion for spectra that we classify as
  diffusive-like. We show that in this case, the spin flip predicted
  along the dispersion is enhanced when the system is close to a
  parametrically amplified instability.
\end{abstract}

\pacs{71.36.+c, 03.75.Mn, 03.75.Kk}





\maketitle

\section{Introduction}
\label{sec:intro}
Strongly coupled matter-light systems, such as exciton-polariton
microcavities, have recently witnessed an escalating interest thanks
to the simultaneous versatility in manipulating and probing their
intrinsic properties. Resulting from the strong coupling of cavity
photons and quantum well excitons, exciton-polaritons display unique
properties deriving from both their constituents --- for recent
reviews see
Refs.~\cite{kavokin_laussy,keeling_review07,keeling_berloff_review,iac_review}. In
particular, the resonant excitation scheme, where polaritons are
directly injected by an external laser near the energy of the lower
polariton dispersion, allows to experimentally access a unique
accurate tuning of the system parameters, such as the polariton
density, current properties, as well as their phase, which is locked
to the one of the external pump.

Much work has been already done both theoretically as well as
experimentally for resonantly pumped single component polariton fluids
in the pump-only
configuration~\cite{carusotto04,ciuti_05,amo09_b,pigeon_11,amo_science11,nardin11,sanvitto11},
i.e., where only the pump state is occupied and no parametric
scattering occurs.
Particular interest was dedicated in analysing the properties of the
collective spectrum of excitations and relate them to the system
superfluid behaviour~\cite{carusotto04,ciuti_05,amo09_b}. Here, the
spectrum could be classified either as gapped, or linear, or else
diffusive-like, in terms of a single parameter, the renormalised pump
detuning.
Interestingly, diffusive-like spectra in non-equilibrium fluids have
been shown to be related to parametric scattering and
amplification~\cite{ciuti_05} and to the occurrence of a negative drag
force of the single component polariton fluid when scattering against
a localised defect~\cite{neg_drag}.

In this article we consider the case of a spinor, i.e., two-component,
polariton fluid, by explicitly including the polarisation degrees of
freedom.
It has been observed that, for fixed pump detuning and degree of
polarisation, the system undergoes a spin flip and a subsequent
hysteresis curve when varying the pump power, promoting this system as
an ideal environment where to realise an optical spin
switch\cite{gippius_multistability, sarkar_mult,nature_polswitch} or a logical gate\cite{kavokin_logic}. 
Interestingly, it has been recently demonstrated that spinor polariton
systems have tunable cross-spin interaction
properties~\cite{takemura2014}.

In previous recent work~\cite{malpuech_disp}, the spectrum of elementary excitation for a resonantly excited pump-only polariton fluid including the spin degrees of freedom was analysed. In that work, by considering for simplicity only the limiting cases of a purely linearly polarised fluid and a purely circularly polarised one, the focus was on the superfluid properties of the system and the possiblity of reproducing a linear spectrum when the pump detuning compensates exactly the interaction-induced blueshift.

Here, we propose a pump-probe set-up tailored for analysing the
properties of the collective excitation spectrum. We show that
in the linear response approximation scheme, valid for a weak probe
beam, the spectrum of excitations can be evaluated analytically even in the generic case of an elliptically polarised spinor polariton fluid. Further, for fixed interaction strengths, the spectrum
can be completely classified in terms of only two experimentally
tunable parameters: the mean-field polarisation angle and a rescaled
pump detuning. Now the number of different class sets of spectra is
much larger compared with the single fluid case. Yet, depending how
the two opposite circular polarisation degrees of freedom mix together
in the spectra, we can single out three larger sets where the
behaviour of the spectrum intrinsic degree of polarisation is
qualitatively different. We name them as \emph{gapped},
$0$-\emph{diffusive}, and $\omega$-\emph{diffusive}; these regions in the two-parameter space are separated by conditions for which the spectrum can be \emph{linear}.
While for gapped spectra, there is no mixing of opposite circular
polarisation degrees of freedom, for both diffusive-like spectra, the
mixing is responsible for flips of the intrinsic spin degree of
polarisation along the branches.
Further, we evaluate the system response to the external probe and
analyse the properties of the transmitted light and its relations to
the collective excitation spectrum. In particular we determine the
properties of the spin flip along the branches for diffusive-like
spectra and how the spin rotation is larger, the closer the system is
to a parametric instability.

The paper is organised as follows: In Sec.~\ref{sec:model}, we present
the generalised Gross-Pitaevskii equation that describes the
resonantly pumped spinor fluid and briefly discuss its mean-field
solutions from existing literature. We introduce the proposed
pump-probe scheme in Sec.~\ref{sec:linea} and discuss the linear
response approximation scheme. The spectrum of collective excitations
is evaluated in Sec.~\ref{sec:spect}, where we derive a ``phase
diagram'' classifying all possible spectral categories. In
Sec.~\ref{sec:eigen} we derive the emission properties of the
intrinsic degree of polarization for each spectrum branch. Finally, in
Sec. \ref{sec:probe}, we evaluate the spinor polariton fluid response
to the additional probe beam and relate its properties to the spectrum
intrinsic properties previously discussed.

\begin{figure}
\centering
\includegraphics[width=3.2in]{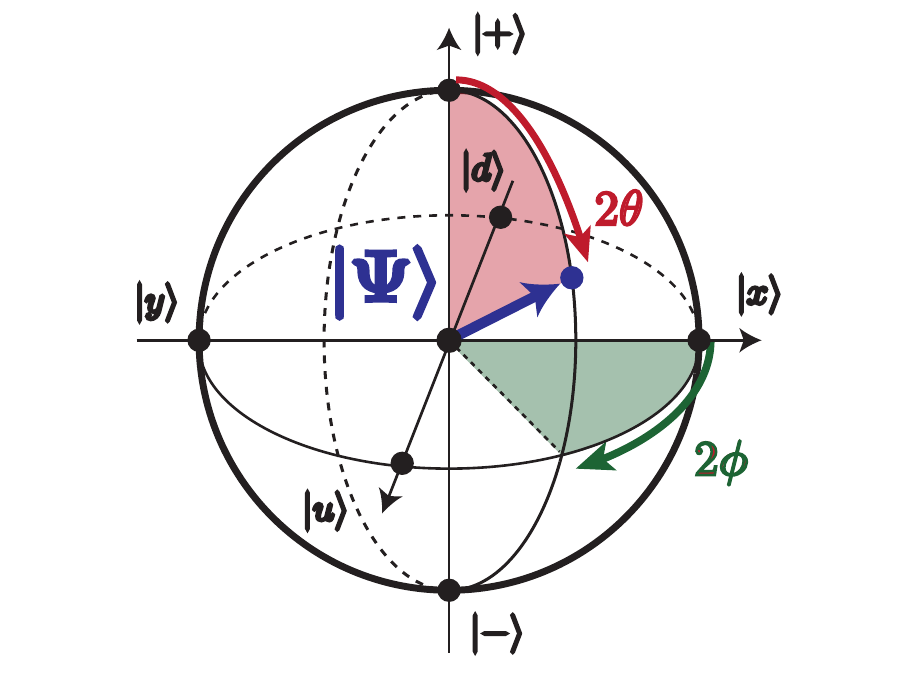}
\caption{(Color online) Representation of the Poincar\'e sphere
  illustrating all possible light polarisations. The basis of left and
  right circular polarisation (perpendicular to the microcavity
  plane), $\{ |+ \rangle, |- \rangle \}$, are the sphere north and
  south poles, respectively. In contrast, all linear polarisation
  states (parallel to the cavity mirror) lie on the equator and for
  those one can choose two alternative basis, either $\{|x\rangle,
  |y\rangle \}$ or $\{|u\rangle, |d\rangle \}$. A generic elliptically
  polarised state $|\Psi \rangle$ lies everywhere on the sphere,
  except at the poles and at the equator, and is defined by a polar
  angle $2 \theta$ (characterising the degree of mixing between
  circular and linear polarisation) and an azimuthal angle $2 \phi$
  (characterising the in-plane polarisation orientation with respect
  to an $x$-linearly-polarised state).}
\label{fig:poinc}
\end{figure}
%
\section{Model}
\label{sec:model}
The dynamics of resonantly pumped polaritons is described by a
Gross-Pitaevskii equation (GPE) for the polariton field generalised to
include the effects of the polariton finite lifetime $2\pi \hbar
/\gamma$, as well as the ones of an external laser that resonantly
injects polaritons into the microcavity~\cite{iac_review}. Here, we
consider a simplified model which involves the lower polariton (LP)
branch only. The resonant pumping scheme we will consider implies
populating a specific LP state with low momentum, allowing us to
neglect the occupation of the upper polariton branch. Further, we
include the two degrees of freedom of the polariton polarisation in
the left $|+\rangle $ and right $|-\rangle $ circular polarisation
basis~\cite{kavokin_laussy,polarization_review} --- see Fig.~\ref{fig:poinc} for a
schematic representation of the Poincar\'e sphere for light
polarisation. The generalised GPE equation for the spinor LP field
$\Psi_{\pm} (\vect{r}, t)$ reads as ($\hbar=1$ throughout):
\begin{multline}
  i \partial_t \Psi_{\pm} = \left[\omega_{LP} (-i \nabla) - i
    \Frac{\gamma}{2} + \alpha_1 |\Psi_{\pm}|^2 + \alpha_2
    |\Psi_{\mp}|^2\right] \Psi_{\pm}\\
  + \mathcal{F_{\pm}} (\vect{r}, t)\; .
\label{eq:model}
\end{multline}
Here, the homogeneous pump term $\mathcal{F_{\pm}} (\vect{r}, t)$,
\begin{equation}
  \mathcal{F_{\pm}} (\vect{r}, t) = f_{\pm}^p e^{i (\vect{k}_p \cdot
    \vect{r} - \omega_p t)} \; ,
\label{eq:pumpo}
\end{equation}
resonantly injects polaritons with a momentum $\vect{k}_p$ and an
energy $\omega_p$ close to the bottom of the LP dispersion
$\omega_{LP} (\vect{k})$. For this reason, we consider a quadratic
approximate of such a dispersion, $\omega_{LP} (\vect{k})
\simeq_{\vect{k} \to 0} \frac{k^2}{2m}$, where $m$ is the LP mass and
we have fixed $\omega_{LP} (0)=0$. In~\eqref{eq:model} we can thus
substitute $\omega_{LP} (-i \nabla) \simeq -\frac{\nabla^2}{2m}$.

Polariton interaction properties depend on their polarisation
component. With $\alpha_1$ we denote the interaction strength between
polaritons in the same circular polarisation state; this is repulsive,
$\alpha_1 > 0$, like the interaction strength between excitons with
the same spin~\cite{Ciuti1998}; when excitons mix with photons to form
polaritons, one can show that the resulting strength $\alpha_1$ weakly
depends on the LP properties such as the photon-exciton detuning and
the Rabi splitting~\cite{vlad_interactions}. Instead $\alpha_2$ is the
interaction strength between polaritons with opposite circular
polarisations. Interestingly, it has been very recently shown that
this inter-polarisation coupling can be tuned by means of a
bipolariton Feshbach resonance mechanism~\cite{takemura2014} from
being attractive to repulsive, by simply changing the value of the LP
photon-exciton detuning. Here, we assume to be far from such a
resonance, in a regime where $\alpha_2$ is weakly attractive, and in
particular we fix $\alpha_2/\alpha_1=
-0.1$~\cite{vlad_interactions}. Note that, for an equilibrium
homogeneous spinor Bose-Einstein condensate at zero temperature (i.e.,
described by the same GPE with no pumping nor decaying term and
chemical potentials fixing the particle number in each condensate),
attraction between opposite components implies a collapse of the
system, i.e., mechanical instability, unless $|\alpha_2|/\alpha_1 <
1$~\cite{pethick2002}. In this regime, the results obtained here for
the polariton spinor fluid resonantly pumped by an external laser do
not qualitatively depend on the particular value chosen for the ratio
$|\alpha_2|/\alpha_1$.  Interestingly, the anisotropy of
polariton-polariton interactions, characterised by the ratio
$\alpha_2/\alpha_1$, was shown to be responsible for the existence of
effective magnetic monopoles in the form of half-integer topological
defects\cite{soln_monopoles} and screening of a magnetic field, the
\emph{spin Meissner effect}\cite{spin_meissner}.

For a homogeneous pump-only scheme as in Eq.~\eqref{eq:pumpo}, the GPE
dynamics~\eqref{eq:model} is solved by the following mean-field
plane-wave steady-state solution,
\begin{equation}
  \Psi_{\pm} (\vect{r}, t) = \psi_{\pm}^{} e^{i (\vect{k}_p \cdot
    \vect{r} - \omega_p t)} \; ,
\label{eq:meanf}
\end{equation}
i.e., by assuming that the pump only populates the LP state with the
very same momentum $\vect{k}_p$ and energy $\omega_p$. Note that in
this resonant pumping scheme, the polariton fluid phase is locked to
the one of the external laser. One can then find how the intensity of
the emission in the two polarisation states, $|\psi_{\pm}^{}|^2$, vary
when changing the system parameters, e.g., by increasing the pump
strength and its degree of polarisation, via the parameters $f_{\pm}$.
Much work has been recently carried out to investigate the
  mean-field properties of spinor polariton fluids, including the
  possibility for multistable behaviour --- see, e.g.,
  Refs.~\cite{wouters_spin_nature, sarkar_mult}. In particular, for a
  single-component resonantly pumped polariton fluid, multistability
  appears when the polariton population goes through an hysteresis
  loop as a function of the pump intensity: When resonantly pumping
  above the LP dispersion, an increase of the pump power implies an
  interaction induced blue-shift of the LP energy towards resonance
  and thus a sudden increase of the LP population. When instead the
  pump power is lowered, a sudden drop occurs at lower pump powers,
  resulting in a region on multistability. For spinor fluids, it has
  been found that the most-populated spin component is subject to an
  hysteresis loop, while the less-populated component undergoes a
  smooth intensity increase~\cite{gippius_multistability,wouters_spin_nature,
    sarkar_mult}. Interestingly, this mechanism has been proposed for
  realising an optical spin switch~\cite{nature_polswitch}. 

We refer the reader to the literature for the mean-field
  analysis, and instead assume here that the system is pumped in such
a way as to induce a given degree of polarisation for the mean-field
solution. This is completely characterised, as any general elliptical
polarisation state, by both a polar angle $\theta_0$ and an azimuthal
angle $\phi_0$ (see Fig.~\ref{fig:poinc}), respectively defined as:
\begin{align}
\label{eq:mfpol}
  \cos(2\theta_0) &= \Frac{|\psi_+^{}|^2 -
    |\psi_-^{}|^2}{|\psi_+^{}|^2 + |\psi_-^{}|^2}\\
  \tan (2\phi_0) &= \Frac{\Im(\psi_+^* \psi_-^{})}{\Re(\psi_+^*
    \psi_-^{})} \; .
\label{eq:mfpo2}
\end{align}
Without loss of generality, we can however assume that the pump
induces a mean-field state with $\phi_0 = 0$, as this simply
corresponds to a choice of the reference coordinate system and any
other elliptically polarised state can be obtained by rotating the
microcavity plane.
We then study how the collective excitation spectrum for such spinor
polariton fluid, as well as its response to a weak-probe beam, change
when varying $\theta_0 \in [0, \pi/4]$ --- the other interval
$\theta_0 \in [\pi/4,\pi/2]$ is symmetric for $\psi_+^{}
\leftrightarrow \psi_-^{}$.

\begin{figure}
\centering
\includegraphics[width=3.2in]{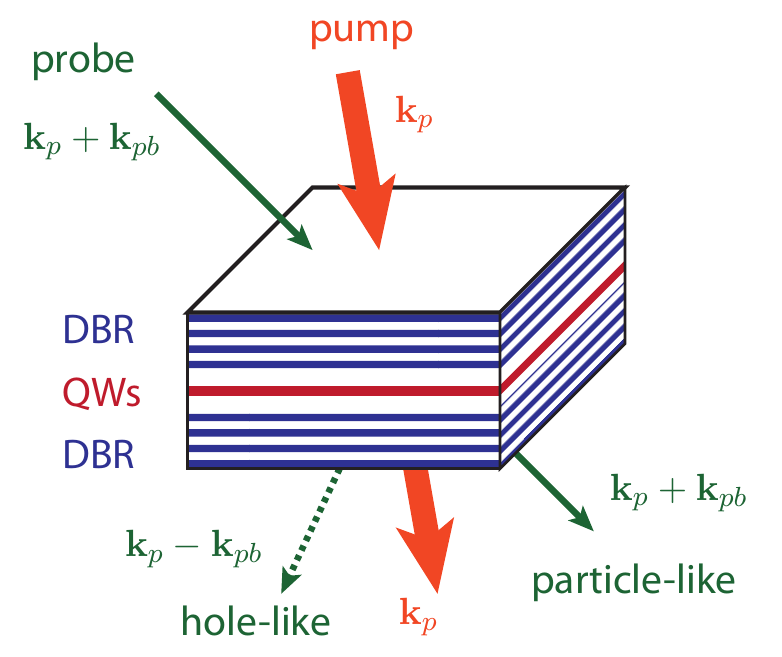}
\caption{(Color online) Schematic setup for the pump-probe experiments
  designed to measure the collective excitation spectrum of a spinor
  polariton fluid. The cavity (composed of two distributed Bragg
  mirrors [DBR] with embedded quantum wells [QWs]) is resonantly
  excited with a pump, with a momentum $\vect{k}_p$ and an energy
  $\omega_p$ close to the LP dispersion. A second weak probe beam,
  with momentum $\vect{k}_p + \vect{k}_{pb}$, and energy
  $\omega_p+\omega_{pb}$ that can be scanned at different values, is
  used to probe the system collective excitation spectrum. The
  transmitted light will have a ``particle-like'' component emitting
  at the direction corresponding to $\vect{k}_p + \vect{k}_{pb}$ and a
  ``hole-like'' component at $\vect{k}_p - \vect{k}_{pb}$.}
\label{fig:schem}
\end{figure}
%
\section{Linear response to a weak probe beam}
\label{sec:linea}
In order to probe the spectrum of collective excitations of a
resonantly pumped spinor polariton fluid, we introduce an additional
weak beam which can be shined at several energies and angles,
different from the pump ones. In particular, referring to the
schematic setup for the proposed pump-probe experiment in
Fig.~\ref{fig:schem}, we consider a homogeneous pump term as
in~\eqref{eq:pumpo} and add to it a homogeneous probe beam with
strength $f_{\pm}^{pb}$, shined at a direction $\vect{k}_p +
\vect{k}_{pb}$ (with $\vect{k}_{pb} \ne 0$) and an energy $\omega_p +
\omega_{pb}$:
\begin{equation}
  \mathcal{F_{\pm}} (\vect{r}, t) = e^{i (\vect{k}_p \cdot \vect{r} -
    \omega_p t)} \left[f_{\pm}^p + f_{\pm}^{pb} e^{i (\vect{k}_{pb}
      \cdot \vect{r} - \omega_{pb} t)} \right]\; .
\label{eq:pumpb}
\end{equation}%
The wave-vector $\vect{k}_{pb}$ should not be confused with the probe
direction, rather it corresponds to the probe momentum relative to the
pump momentum $\vect{k}_p$.

We assume that the system is only weakly perturbed by the probe;
therefore, we can apply a linear-response approximation, where only
two other states are weakly populated aside the mean-field
state~\eqref{eq:meanf}~\cite{michiel_probe}:
\begin{multline}
  \Psi_{\pm} (\vect{r}, t) = e^{i (\vect{k}_p \cdot \vect{r} -
    \omega_p t)} \left[\psi_{\pm}^{} \phantom{u_{\pm}^{} e^{i
        (\vect{k}_{pb} \cdot \vect{r} - \omega_{pb} t)}} \right. \\
  \left. + u_{\pm}^{} e^{i (\vect{k}_{pb} \cdot \vect{r} - \omega_{pb}
    t)} + v_{\pm}^* e^{-i (\vect{k}_{pb} \cdot \vect{r} - \omega_{pb}
    t)} \right]\; .
\label{eq:fullr}
\end{multline}
Note that, although the polariton sample is only excited at two
directions, the pump $\vect{k}_p$ and the probe one $\vect{k}_p +
\vect{k}_{pb}$, transmission must also include an additional signal at
$\vect{k}_p - \vect{k}_{pb}$. This is a consequence of polariton
interactions which mix the particle-like excitations $u_{\pm}^{}$,
resulting from adding a particle into the mean-field state, with the
hole-like degrees of freedom $v_{\pm}^{}$, which are excited by
instead removing a particle.
Thus, as schematically drawn in Fig.~\ref{fig:schem}, we expect
  the weak probe to imply a transmission in both directions
  $\vect{k}_p \pm \vect{k}_{pb}$.  We will analyse in
  Sec.~\ref{sec:probe} the properties of both transmission signals, as
  well as the relation to the intrinsic properties of the collective
  excitation spectrum.

The system response to the probe is easily evaluated by
substituting~\eqref{eq:fullr} into the GPE equation~\eqref{eq:model}
and by expanding at first order in both the probing field strength
$f_{\pm}^{pb}$ and the fluctuation terms above mean-field, $u_{\pm}$
and $v_{\pm}$. 

We obtain four coupled equations diagonal in momentum
space
\begin{equation}
  \left(\omega_{pb}\hat{\mathbb{I}} -
  \hat{\mathcal{L}}_{\vect{k}_{pb}}\right) \vect{w} = \vect{f}^{pb} \;
  ,
\label{eq:invre}
\end{equation}
where response and probe have been rearranged into four component
vectors, $\vect{w} = (u_+^{}, v_+^{}, u_-^{}, v_-^{})^T$ and
$\vect{f}^{pb} = (f_{+}^{pb}, 0, f_{-}^{pb}, 0)^T$. The Bogoliubov
operator $\hat{\mathcal{L}}_{\vect{k}}$ can be written in terms of its
polarisation components,
\begin{equation}
  \hat{\mathcal{L}}_{\vect{k}} = \begin{pmatrix}
    \hat{\mathcal{M}}_{++,\vect{k}} & \hat{\mathcal{M}}_{+-,\vect{k}}
    \\ \hat{\mathcal{M}}_{-+,\vect{k}} &
    \hat{\mathcal{M}}_{--,\vect{k}}\end{pmatrix} \; ,
\label{eq:bogol}
\end{equation}
which are given by the expressions ($i=+, -$):
\begin{align}
  \hat{\mathcal{M}}_{ii,\vect{k}} &= \begin{pmatrix}
    \tilde{\epsilon}_{i,\vect{k}} + \vect{k} \cdot \vect{v}_p - i
    \frac{\gamma}{2} & \alpha_1 {\psi_i^{}}^2 \\ -\alpha_1
         {\psi_i^*}^2 & -\tilde{\epsilon}_{i,\vect{k}} + \vect{k}
         \cdot \vect{v}_p - i \frac{\gamma}{2} \end{pmatrix}\\
  \hat{\mathcal{M}}_{+-,\vect{k}} &= \alpha_2 \begin{pmatrix} \psi_+^*
    \psi_-^{} & \psi_+^{} \psi_-^{} \\ - \psi_+^* \psi_-^* &
    -\psi_+^{} \psi_-^* \end{pmatrix}\; .
\end{align}
The parameters appearing in the diagonal components of the Bogoliubov
operator are the fluid velocity $\vect{v}_p = \vect{k}_p/m$ and the
following energy term:
\begin{align}
  \tilde{\epsilon}_{\pm, \vect{k}} &= \epsilon_{\pm, \vect{k}} +
  \alpha_1 |\psi_{\pm}^{}|^2 \\
  \epsilon_{\pm, \vect{k}} &= \Frac{\vect{k}^2}{2m} - \Delta_{\pm} \\
  \Delta_{\pm} &= \omega_p - \left(\Frac{\vect{k}_p^2}{2m} + \alpha_1
    |\psi_{\pm}^{}|^2 + \alpha_2 |\psi_{\mp}^{}|^2\right) \; \label{eq:pdetuning}.
\end{align}
In particular, $\Delta_{\pm}$ can be interpreted as the effective pump
detuning, i.e., the energy difference between the laser frequency
$\omega_p$ and the LP dispersion at momentum $\vect{k}_p$ renormalised
by the interaction induced blue-shift due to both the
intra-polarisation coupling $\alpha_1 |\psi_{\pm}^{}|^2$ and the
inter-polarisation one $\alpha_2 |\psi_{\mp}^{}|^2$.

Before analysing the properties of the probe response $\vect{w}$
starting from Eq.~\eqref{eq:invre} (Sec.~\ref{sec:probe}), we discuss
first the collective excitation spectrum of the spinor polariton fluid
and its intrinsic properties (Sec.~\ref{sec:spect}), including its
degree of polarisation (Sec.~\ref{sec:eigen}).

\subsection{Excitation spectrum of  the spinor polariton fluid}
\label{sec:spect}
For a general wave-vector $\vect{k}$ (as for $\vect{k}_{pb}$, here
$\vect{k}$ is assumed to be measured with respect to the pump
wave-vector $\vect{k}_p$), the different branches of the spectrum of
excitations are the eigenvalues of the Bogoliubov operator
$\hat{\mathcal{L}_{\vect{k}}}$ and thus are evaluated starting from
the equation:
\begin{equation}
  \det \left(\hat{\mathcal{L}_{\vect{k}}} - 
    \omega \hat{\mathbb{I}}\right) = 0\; ,
\end{equation}
or equivalently finding the roots of the following polynomial
equation:
\begin{multline*}
  \prod_{i=+, -} \left[\left(\omega + i \frac{\gamma}{2} - \vect{k}
      \cdot \vect{v}_p\right)^2 - E_{i, \vect{k}}^2\right] \\
  = - 4 \alpha_2^2 \prod_{i=+, -} \left(\epsilon_{i, \vect{k}}
    |\psi_{i}^{}|^2 \right)\; .
\end{multline*}
This can be solved exactly, resulting in four branches of the
spectrum, which, as explained later, we label with a new index $a
  = u_\uparrow, v_\uparrow, u_\downarrow, v_\downarrow$:
\begin{multline}
  \omega_{\vect{k}}^{(a)} = \vect{k} \cdot \vect{v}_p -
  i\frac{\gamma}{2} + \eta_{u,v} \left[\Frac{E_{+, \vect{k}}^2 + E_{-,
        \vect{k}}^2}{2} \right. \\
  \left. +\sigma_{\uparrow,\downarrow} \sqrt{\left(\Frac{E_{+,
        \vect{k}}^2 - E_{-, \vect{k}}^2}{2}\right)^2 + 4 \alpha_2^2
    \prod_{i=+, -} \left(\epsilon_{i, \vect{k}} |\psi_{i}^{}|^2
    \right)} \right]^{1/2}
\label{eq:collsp}
\end{multline}
where $\eta_{u,v}=\pm 1$ for the particle-like and hole-like
  components respectively and $\sigma_{\uparrow,\downarrow}= \pm 1$.
Here, the energy
\begin{equation}
  E_{\pm, \vect{k}} = \sqrt{\epsilon_{\pm, \vect{k}} (\epsilon_{\pm,
      \vect{k}} + 2 \alpha_1 |\psi_{\pm}^{}|^2)} \; ,
\end{equation}
determines the excitation spectrum of two independent fluids with
opposite circular polarisations, which is given
by~\cite{carusotto04,andrei_drag}
\begin{equation}
  \lim_{\alpha_2 \to 0} \omega_\vect{k}^{(a)} = 
  \vect{k} \cdot \vect{v}_p - i\frac{\gamma}{2} + \eta_{u,v} 
  E_{\pm, \vect{k}} \; .
\label{eq:indpo}
\end{equation}
When setting $\psi_-=0$ (circular polarization) or $\psi_+=\psi_-$ (linear polarization),  one recovers the limiting expressions derived in Ref. \cite{malpuech_disp}.

Note that, because of interactions, polarisation and particle-hole
degrees of freedom do in general mix together along the dispersion of
each spectrum branch. Yet, the choice of the index $a = u_\uparrow,
v_\uparrow, u_\downarrow, v_\downarrow$ for labeling the four branches
of the excitation spectrum is motivated by the fact that, at large
momenta there is no mixing between the particle-like
($u_{\uparrow,\downarrow}$) and hole-like ($v_{\uparrow,\downarrow}$)
degrees of freedom, while the same does not hold of $+$ and $-$
polarisation states that do remain coupled --- i.e., as we will see
later, the intrinsic polarisation of these branches can never be
purely circularly $+$ or $-$ polarised even at large momenta, where
the energy becomes
\begin{multline}
  \lim_{k \to \infty} \omega_\vect{k}^{(u_\uparrow, v_\uparrow,
    u_\downarrow, v_\downarrow)} = \vect{k} \cdot \vect{v}_p -
  i\Frac{\gamma}{2}\\ 
  +\eta_{u,v} \left[\Frac{(\vect{k}+\eta_{u,v}\vect{k}_p)^2}{2m} +2
    \alpha_1 |\psi_{\uparrow, \downarrow}^{}|^2 + \alpha_2
    |\psi_{\downarrow, \uparrow}^{}|^2\right]\; ,
\label{eq:freed}
\end{multline}
where $\psi_{\uparrow, \downarrow}^{} \equiv \psi_{+,-}^{}$. For this
reason, we introduces a new notation $\uparrow,\downarrow$ for the
branch index $a= (u_\uparrow, v_\uparrow, u_\downarrow,
v_\downarrow)$, indicating that the pure circular polarisation degrees
of freedom $\pm$ are always coupled.

\begin{figure}
\centering
\includegraphics[width=3.2in]{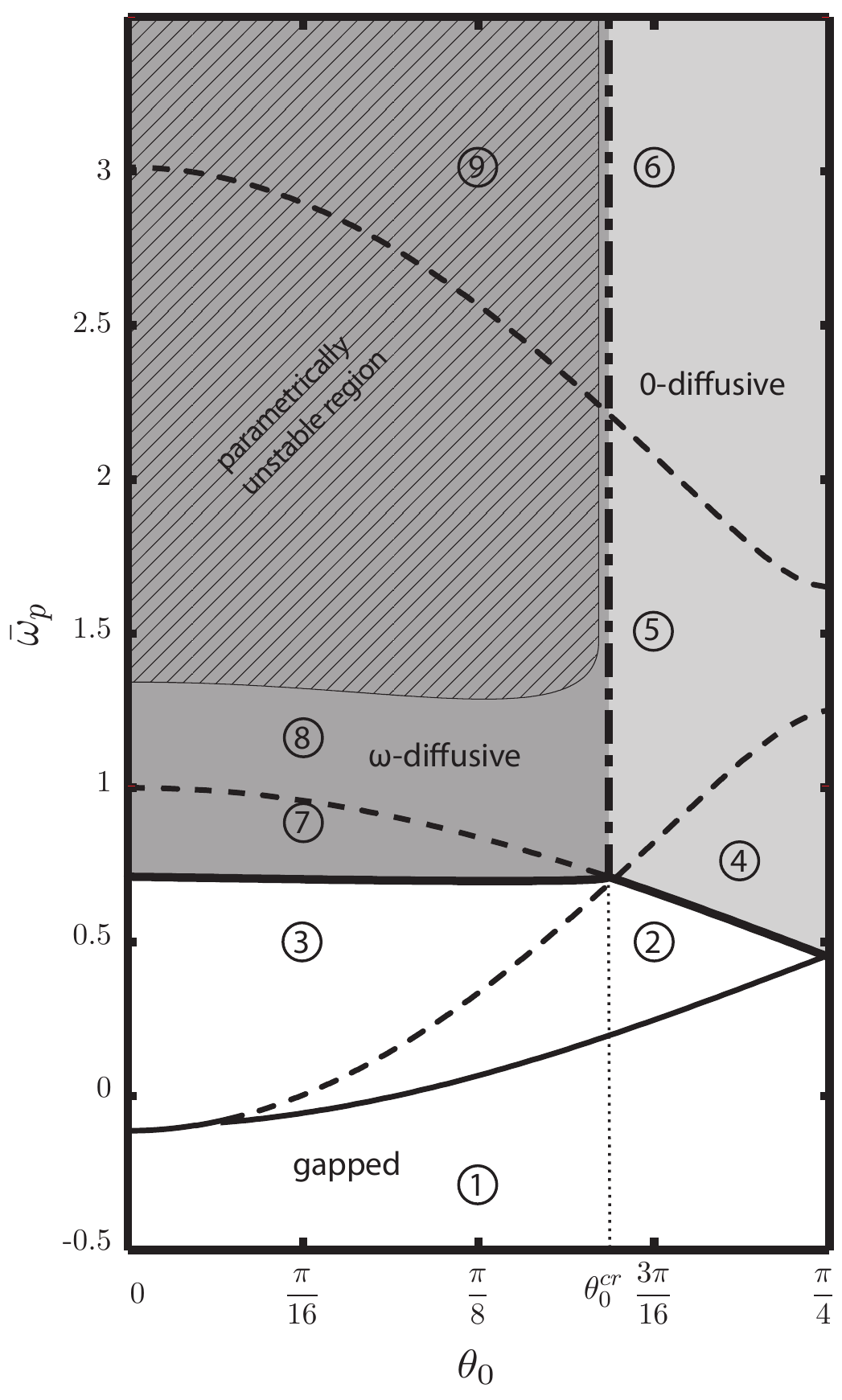}
\caption{``Phase diagram'' showing the different classes of spectra of
  a spinor polariton fluid as a function of the two dimensionless
  parameters $\bar{\omega}_p$~\eqref{eq:resen} and the mean-field
  polarisation angle $\theta_0$~\eqref{eq:mfpol} for
  $\alpha_2/\alpha_1 = -0.1 $--- the case of left circular
  polarisation corresponds to $\theta_0 = 0$, while the one of linear
  polarisation to $\theta_0 = \pi/4$. The labels $1-9$ correspond to
  the exact parameter values chosen for the corresponding spectra
  plotted in Fig.~\ref{fig:spect}. The value of the critical
  polarisation angle $\theta_0^{\text{cr}}$~\eqref{eq:thcri} is marked
  with a dash-dotted line. The white region (\emph{gapped}) includes
  the spectra $1-3$ characterised by a gap for the $+$ branches; the
  clear-gray region ($0$-\emph{diffusive}) are the spectra $4-6$
  displaying diffusive behaviour at zero energy, while the dark-gray
  region ($\omega$-\emph{diffusive}) are the spectra $7-9$ where a
  diffusive region can also be at finite energy (see text). The
  striped region is the parametrically unstable region for
  $\gamma=1.5\mathcal{E}$ .}
\label{fig:phdia}
\end{figure}
\begin{figure}
\centering
\includegraphics[width=3.2in]{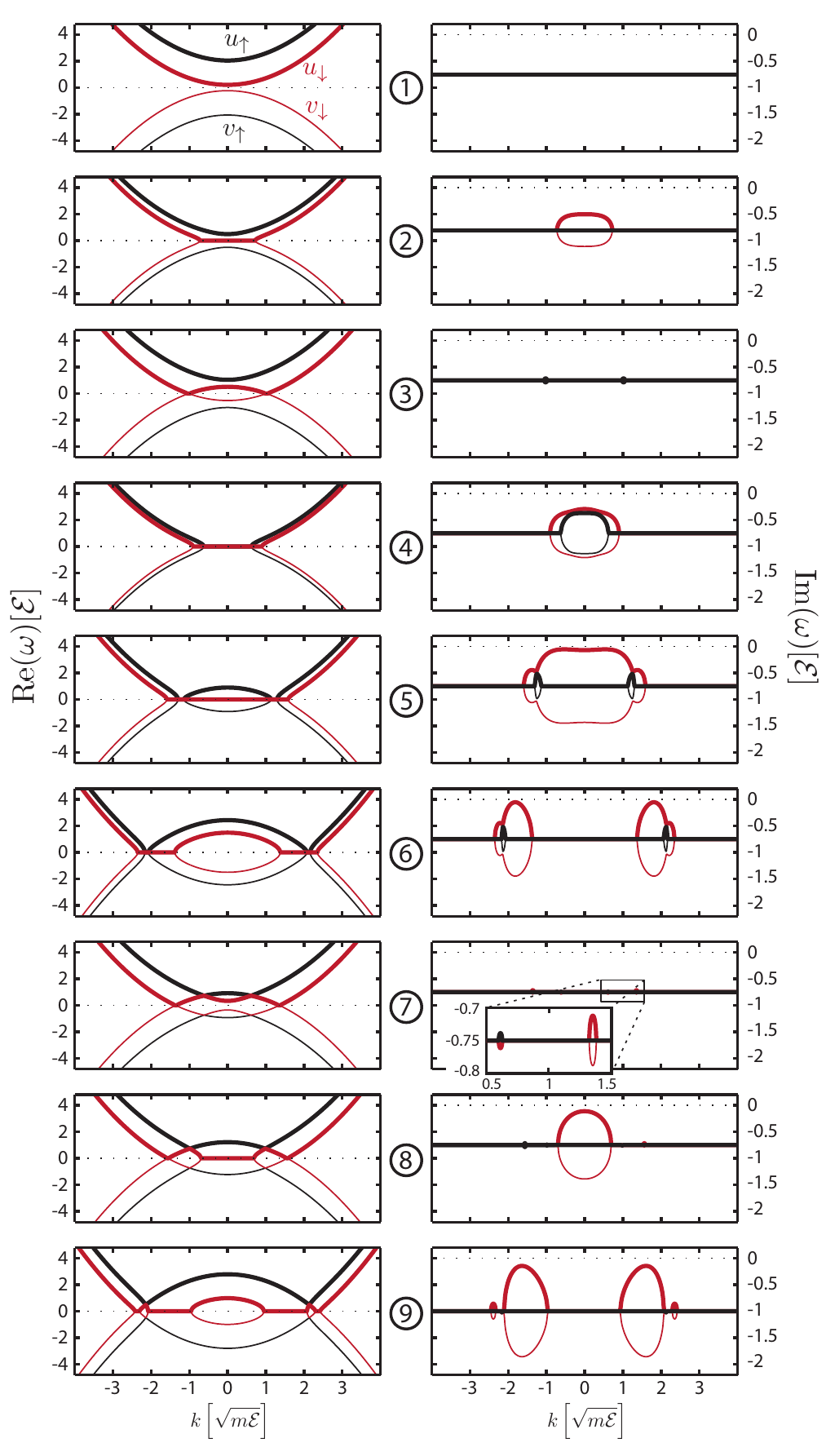}
\caption{(Color online) Different types of excitation spectra (real
  part $\Re \omega^{(a)}_{\vect{k}}$ in the left panels and imaginary
  part $\Im \omega^{(a)}_{\vect{k}}$ in the right panels) allowed for
  a spinor polariton fluid for $\alpha_2/\alpha_1 = -0.1$ and for a
  pump wave-vector $\vect{k}_p=0$. The energy $\omega$ is measured in
  units of $\mathcal{E} = \alpha_1 (|\psi_+^{}|^2 + |\psi_-^{}|^2)$,
  while momentum (plots are cuts at $k_y=0$) in units of $\sqrt{m
    \mathcal{E}}$. Thick (thin) black lines are the $u_{\uparrow}$
  ($v_{\uparrow}$) branches, while thick (thin) red-gray lines are the
  $u_{\downarrow}$ ($v_{\downarrow}$) branches. The labels $1-9$
  correspond to the very same parameters $(\bar{\omega}_p , \theta_0)$
  shown in the ``phase diagram'' of Fig.~\ref{fig:phdia}, where the
  corresponding labels appear. The polariton decay rate is fixed to
  $\gamma = 1.5\mathcal{E}$ for the spectra $1-8$ and to
  $\gamma=2\mathcal{E}$ for spectrum $9$.}
\label{fig:spect}
\end{figure}
We now classify all possible different types of excitation spectra
(see Fig.~\ref{fig:spect}) and how these evolve from one type to the
other by changing the system parameters, as represented in the ``phase
diagram'' of Fig.~\ref{fig:phdia}.
Interestingly, for a fixed interaction strength ratio
$\alpha_2/\alpha_1$ ($=-0.1$ in the figures), only two dimensionless
independent parameters are sufficient in order to classify all
possible different types of spectra of a resonantly pumped spinor
polariton fluid: 1) the mean-field polarisation angle
$\theta_0$~\eqref{eq:mfpol} and 2) the dimensionless pump energy
rescaled by the ``self-interaction'' energy $\mathcal{E} = \alpha_1
(|\psi_+^{}|^2 + |\psi_-^{}|^2)$:
\begin{equation}
  \bar{\omega}_p = \Frac{\omega_p}{ \alpha_1 (|\psi_+^{}|^2 +
    |\psi_-^{}|^2)} \; .
\label{eq:resen}
\end{equation}

By means of these two parameters only we can fully classify all the
allowed spectrum typologies. Note in fact that the value of the pump
momentum $\vect{k}_p$ has the sole effect of tilting the spectrum
dispersion; in the quadratic approximation for the LP dispersion
considered here, this corresponds to a Galilean
transformation.~\cite{landau_lifshitz_FD} For this reason, we plot,
without loss of generality, the spectra of Fig.~\ref{fig:spect} for a
pump in the orthogonal direction to the cavity growth, $\vect{k}_p =
0$.

In the absence of inter-polarisation interaction, $\alpha_2 = 0$, and
for equal spin populations, $|\psi_+^{}| = |\psi_-^{}|$, both sign and
value of a single parameter, the rescaled interaction renormalised
pump detuning $\bar{\Delta} = \Delta/\alpha_1 |\psi_+|^2$, determine
the four types of possible spectra~\cite{carusotto04, andrei_drag}:
1) for $\Delta < 0$ the spectrum is \emph{gapped}; 2) the gap closes
to zero for $\Delta= 0$ and the dispersion is \emph{linear} at low
momenta; while for $\Delta > 0$, particle and hole branches of the
spectrum real part touch together in either 3) one ($\bar{\Delta} \le
2$) or 4) two ($\bar{\Delta} > 2$) separate momentum intervals ---
note that Fig.~\ref{fig:spect} is a cut at $k_y=0$, so intervals for
those plots in reality corresponds to rings in the two-dimensional
$\vect{k}$-space. Both cases 3) and 4) are generally named as
\emph{diffusive} spectra. Note that the linear spectrum is allowed for
a single value of the detuning $\Delta$, and thus even if the types of
different spectra for $\alpha_2 = 0$ are four in total, the finite
interval regions in $\Delta$ displaying different spectra are only
three (and $\Delta= 0$ represents a separating point between two of
these regions). For the spinor case, the minimal set of independent
dimensionless parameters characterising the spectrum is instead formed
by $\theta_0$ and $\bar{\omega}_p$. Note that by rescaling the pump
detuning $\Delta_{\pm}$~\eqref{eq:pdetuning} by the self-interaction
energy $\mathcal{E}$ would still lead to a parameter depending on
$\theta_0$.

For a coupled spinor fluid, with $\alpha_2 \neq 0$, the classes of
different spectra increase from four to eighteen; by counting the
parameter finite regions only (and excluding the separating lines),
this corresponds to nine regions of different spectra compared to the
three of the previous $\alpha_2 \to 0$ limit case.
The proliferation of different types of spectra is due to the presence
of two nested square roots in Eq.~\eqref{eq:collsp}. Which of the four
branches have a degenerate real part as well as in how may momentum
intervals degeneracy occurs, depends on the sign of both square root
arguments. All nine possibilities for the spectra are plotted in
Fig.~\ref{fig:spect} and the various ``phase diagram'' regions in the
$(\bar{\omega}_p, \theta_0)$ parameter space where such spectra are
allowed are plotted in Fig.~\ref{fig:phdia}.

For both negative as well as small positive values of the renormalised
dimensionless pump energy detuning, $\bar{\omega}_p$, the spectrum is
fully \emph{gapped}, i.e., none of the four branches mix together
(panel $1$ in Fig.~\ref{fig:spect}). By increasing the value of
$\bar{\omega}_p$ at fixed $\theta_0$, the $\uparrow$ branches (black
lines) are still gapped, while the real part of the particle-like
$u_{\downarrow}$ (thick red-gray) and hole-like $v_{\downarrow}$ (thin
red-gray) branches undergo the same changes as previously described
for a single component fluid: they touch each other first in a single
momentum interval (panel $2$) and then in two separate momentum
intervals (panel $3$).

Even if for panels $2$ and $3$ the $\downarrow$ branches have a
diffusive-like character, with an imaginary part deviating from the
polariton lifetime, $\Im \omega \ne -\gamma/2$, in all the three
regimes $1-3$ described, the $\uparrow$ and $\downarrow$ real part
branches are never degenerate and hence do not mix one with the other,
so that each maintains its own character: we group the three cases as
\emph{gapped} spectra (white region of Fig.~\ref{fig:phdia}).

When we further increase the value of $\bar{\omega}_p$, the opposite
polarisation branches can however mix together and the spectrum
evolves differently depending on the value of the mean-field polar
angle $\theta_0$: in particular it either changes from the type $3$ to
$7$ if $\theta_0 < \theta_0^{\text{cr}}$, or from $2$ to $4$ if
$\theta_0 > \theta_0^{\text{cr}}$, where the critical angle
$\theta_0^{\text{cr}}$ is given by:
\begin{equation}
  \cos(2\theta_0^{\text{cr}}) = \Frac{1}{2} \sqrt{1 +
    \Frac{\alpha_2}{\alpha_1}} \; .
\label{eq:thcri}
\end{equation}
For $\alpha_2/\alpha_1 = -0.1$, this critical value of the
polarisation angle is given by $\theta_0^{\text{cr}} \simeq 0.54$. The
dash-dotted line in Fig.~\ref{fig:phdia} indicates where
$\theta_0^{\text{cr}}$ determines the boundary between these two
cases.
The difference between the two is how the $\uparrow$ and $\downarrow$
branches mix together and at which energy the mixing happens. For
$\theta_0 > \theta_0^{\text{cr}}$, there is only mixing of the
branches at zero energy, as for the three spectra $4-6$, which we
group under the naming $0$-\emph{diffusive} (clear gray region of
Fig.~\ref{fig:phdia}).  When $\theta_0 < \theta_0^{\text{cr}}$, the
spectra are characterized by the presence of a diffusive momentum
region at finite energy, as for the spectra $7-9$, that we group under
the naming $\omega$-\emph{diffusive} spectra (dark gray region of
Fig.~\ref{fig:phdia}).

If we increase $\bar{\omega}_p$ for $\theta_0>\theta_0^{\text{cr}}$,
the branches transforms via different $0$-\emph{diffusive} phases. For
the spectrum of type $4$ all four branches real parts are degenerate
at $\omega=0$ around $\vect{k}=0$, this then evolves to a zero energy
degeneracy in different parts of the momentum space for the cases $5$
and $6$.
When $\bar{\omega}_p$ is instead increased for
$\theta_0<\theta_0^\text{cr}$, the $\uparrow$ branch transforms via
different $\omega$-\emph{diffusive} phases: Here, we get a mixing of
the $u_\uparrow$ ($v_\uparrow$) branch with the $u_\downarrow$
($v_\downarrow$) branch as in the panel $7$, where there is a narrow
region in $\vect{k}$ space where the spectrum is diffusive and
$\Re\omega \neq 0$. Note that for these spectral types the four
branches cannot be degenerate all at the same time as in the
$0$-\emph{diffusive} case previously considered; now only two branches
at the time get degenerate, while the other two do repel each other.
Higher values of $\bar{\omega}_p$ induce a similar behaviour but for
different intervals of momentum, as for region $8$ or region
$9$. Degeneracy at zero energy is however still possible for the
$\downarrow$ branch, as seen in panels $8$ and $9$.

The spectral phases of a circularly (linearly) polarized fluid are the ones found on the line $\theta_0 = 0$ ($\theta_0=\pi/4$) and coincide with the results discussed in Ref. \cite{malpuech_disp}. 
After having classified completely all possible excitation spectra, we
now discuss in the next section their intrinsic polarisation
properties.

\begin{figure}
\centering
\includegraphics[width=3.2in]{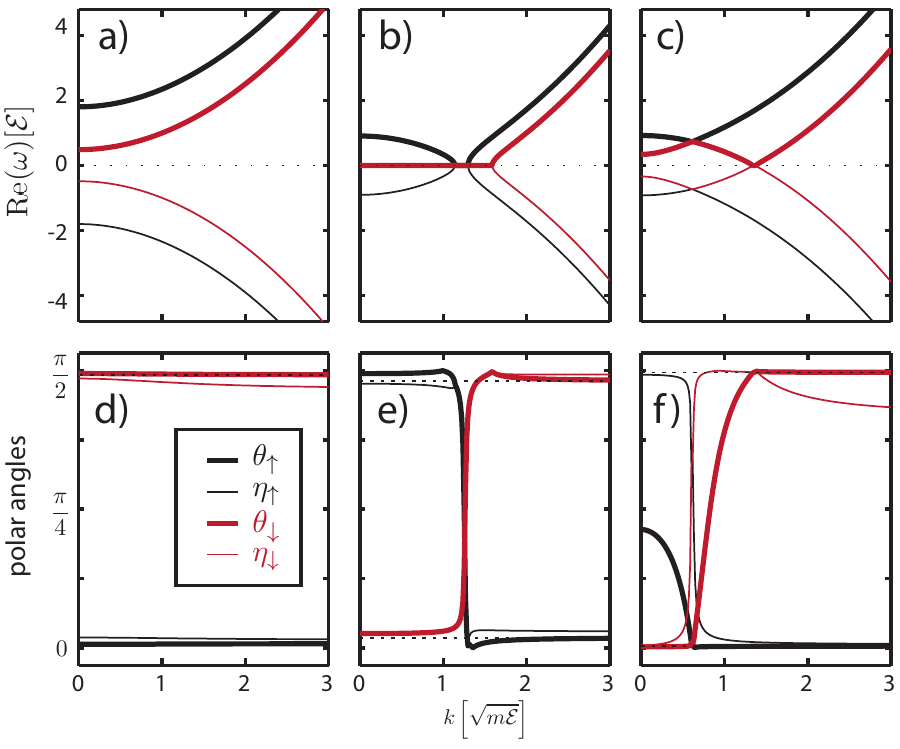}
\caption{(Color online) Bottom panels: Normal $\theta_{\uparrow,
    \vect{k}}$ (thick black lines) and $\theta_{\downarrow, \vect{k}}$
  (thick red-gray) and antinormal $\eta_{\uparrow, \vect{k}}$ (thin
  black) and $\eta_{\downarrow, \vect{k}}$ (thin red-gray) polar
  angles for the \emph{gapped} spectrum $1$ (left panel), the
  $0$-\emph{diffusive} spectrum $5$ (middle panel), and the
  $\omega$-\emph{diffusive} spectrum $7$ (right panel) of
  Fig.~\ref{fig:spect}. For immediate comparison, each spectrum is
  re-plotted in the corresponding upper panels. The asymptotic large
  momentum behaviors for the normal angles~\eqref{eq:limit} are
  plotted as black dashed line. The system parameters are
  $\alpha_2/\alpha_1=-0.1$, $\vect{k}_p=0$, $\gamma=1.5$, and values
  of $(\bar{\omega}_p,\theta_0)$ as specified by the labels $1$, $5$
  and $7$ in the phase diagram of Fig.~\ref{fig:phdia}.}
\label{fig:eigen}
\end{figure}
%
\subsection{Degree of polarisation of collective excitations}
\label{sec:eigen}
Each mode of the collective spectrum does emit with an intrinsic
degree of polarisation along the dispersion. This can be determined by
starting from the eigenvalue equations,
\begin{equation}
  \hat{\mathcal{L}}_{\vect{k}} \vect{x}_{\vect{k}}^{(a)} =
  \omega_{\vect{k}}^{(a)} \vect{x}_{\vect{k}}^{(a)}\; ,
\end{equation}
where $\omega_\vect{k}^{(a)}$ are the four branches~\eqref{eq:collsp}
labeled by the index $a = u_\uparrow, v_\uparrow, u_\downarrow,
v_\downarrow$ and $\vect{x}^{(a)}_\vect{k}$ are the four
four-component eigenvectors of the Bogoliubov matrix
$\hat{\mathcal{L}}_{\vect{k}}$, $\vect{x}^{(a)}_{\vect{k}} = (x_{u_+,
  \vect{k}}^{(a)}, x_{v_+, \vect{k}}^{(a)}, x_{u_-, \vect{k}}^{(a)},
x_{v_-, \vect{k}}^{(a)})$ --- thus here the lower indices $u_+, v_+,
u_-, v_-$ specify the eigenvector component, while the upper indices
$(a)$ refer to the eigenvalue branch.

For each branch labeled by $(a)$ and for a given direction
  $\vect{k}$ (measured with respect to the pump wave-vector
  $\vect{k}_p$) we might expect that, out of a $4$-component complex
  vector $\vect{x}^{(a)}_{\vect{k}}$, the degree of polarisation will
  be characterised by two polar angles and by two azimuthal angles,
  all independent from each other. However, the particle-hole symmetry
  which characterises the Bogoliubov matrix~\eqref{eq:bogol} leads to a
  redundancy of parameters. In particular, the Bogoliubov
  matrix~\eqref{eq:bogol} is symmetric under the simultaneous exchange
  of the $u \leftrightarrow v$ components and the transformation
  $\hat{\mathcal{L}}_{\vect{k}} =
  -\hat{{\mathcal{L}}}_{-\vect{k}}^*$. For eigenvalues and
  eigenvectors this implies that
\begin{align*}
  \omega_{-\vect{k}}^{(u_j)} &= - {\omega_{ \vect{k}}^{(v_j)}}^* \\
  x_{u_\pm,\vect{k}}^{(u_j)} &= - {x_{v_\pm,\vect{k}}^{(v_j)}}^*\; ,
\end{align*}
where $j=\uparrow,\downarrow$.
Thus, in order to characterise the intrinsic degree of polarisation of
the collective spectrum, it is enough to define a ``normal'' polar
polarisation angles for each of the two $j=\uparrow,\downarrow$
particle-like branches as
\begin{equation}
  \cos (2 \theta_{j, \vect{k}}) = |x_{u_+,\vect{k}}^{(u_j )}|^2 -
  |x_{u_-, \vect{k}}^{(u_j)}|^2 \; .
\end{equation}
In fact, for the hole-like branches, we have that $\cos (2 \theta_{j,
  \vect{k}}) = |x_{v_+,-\vect{k}}^{(v_j )}|^2 -
|x_{v_-,-\vect{k}}^{(v_j)}|^2$ --- we are assuming here the following
normalisation conditions for the eigenvector components,
$|x_{u_+,\vect{k}}^{(u_j )}|^2 + |x_{u_-, \vect{k}}^{(u_j)}|^2 = 1 =
|x_{v_+,-\vect{k}}^{(v_j )}|^2 + |x_{v_-,-\vect{k}}^{(v_j)}|^2 $.
Similarly, we can also define two ``anti-normal'' polar polarisation
angles as:
\begin{equation}
  I_{j,\vect{k}} \cos(2 \eta_{j, \vect{k}}) = |x_{u_+,\vect{k}}^{(v_j
    )}|^2 - |x_{u_-, \vect{k}}^{(v_j)}|^2 \; ,
\end{equation}
where $I_{j,\vect{k}} = |x_{u_+,\vect{k}}^{(v_j )}|^2 + |x_{u_-,
  \vect{k}}^{(v_j)}|^2 = |x_{v_+,-\vect{k}}^{(u_j)}|^2 +
|x_{v_-,-\vect{k}}^{(u_j)}|^2$ is the normalisation of the anti-normal
modes with respect to the normalisation of the normal ones which was
fixed to $1$. 
Thanks to the particle-hole symmetry, we could have equivalently
defined these angles as $I_{j} \cos(2 \eta_{j, \vect{k}}) =
|x_{v_+,-\vect{k}}^{(u_j )}|^2 - |x_{v_-,-\vect{k}}^{(u_j)}|^2$.
To summarise, for each branch we have two polar angles $\theta$ and
$\eta$ determining the degree of polarisation of the collective
emission. These are not directly measurable quantities.  However, in
the pump-probe experiment proposed and analysed in the next section,
the normal angle $\theta$ is related to the resonant transmission of
the particle modes along the particle-like probe direction
$\vect{k}_p+\vect{k}_{pb}$, while the anti-normal angle $\eta$ to the
transmission along the hole-like direction $\vect{k}_p-\vect{k}_{pb}$.

We now analyse in the bottom panels of Fig.~\ref{fig:eigen}, the
behaviour of both $\theta_{j, \vect{k}}$ (thick lines) and $\eta_{j,
  \vect{k}}$ (thin lines) along each branch dispersion for three
representative spectra (top panels) of the \emph{gapped},
$0$-\emph{diffusive}, and $\omega$-\emph{diffusive} types previously
classified in Figs.~\ref{fig:phdia} and~\ref{fig:spect}.
Let us first note that it can be easily shown that the intensity
$I_{j,\vect{k}}$ decays quickly to zero at large momenta,
i.e. $I_{j,\vect{k}} \sim k^{-4}$ for $k \gg \sqrt{m \mathcal{E}}$,
where $\mathcal{E} = \alpha_1 (|\psi_+^{}|^2 + |\psi_-^{}|^2)$. This
implies that, in this limit, the coupling between particle-like and
hole-like degrees of freedom can be neglected, allowing us to find the
asymptotic behaviour of the normal polarisation angles at large
momenta:
\begin{align}
\label{eq:limit}
  \lim_{k \gg \sqrt{m \mathcal{E}}} \cos (2 \theta_{j, \vect{k}}) &=
  \Frac{1 + \sigma_j \sqrt{1 + \xi^2}}{1 + \xi^2 +\sigma_j \sqrt{1 +
      \xi^2}}\\
  \xi &= \Frac{\alpha_2}{2 \alpha_1 - \alpha_2} \tan (2 \theta_0)\; ,
  \nonumber
\end{align}
where $\sigma_{j=\uparrow,\downarrow}=\pm 1$ and $\tan (2 \theta_0) =
2 |\psi_+^{}| |\psi_-^{}| /(|\psi_+^{}|^2 - |\psi_-^{}|^2)$.

As panel (d) of Fig.~\ref{fig:eigen} shows, for a \emph{gapped}
spectrum (panel (a), corresponding to spectrum 1 in
Fig. \ref{fig:spect}) the angles $\theta_{\uparrow, \vect{k}}$ (thick
black line) and $\theta_{\downarrow, \vect{k}}$ (thick red-gray line)
vary only very little along the dispersion because there is no mixing
between $\uparrow$ and $\downarrow$ branches. In particular, while for
the $\uparrow$ branch the normal degree of polarisation (thick black)
is almost everywhere fully left-polarised, $\theta_{\uparrow,
  \vect{k}} \simeq 0$ (corresponding to the north pole of the
Poincar\'e sphere in Fig.~\ref{fig:poinc}), for the $\downarrow$
branch (thick red-gray line) $\theta_{\downarrow, \vect{k}} \simeq
\pi/2$ (south pole). Both anti-normal angles $\eta_{j, \vect{k}}$
(thin lines) also display a small variation along the dispersion from
the values $\eta_{\uparrow, \vect{k}} \simeq 0$ (thin black) and
$\eta_{\downarrow, \vect{k}} \simeq \pi/2$ (thin red-gray).

Mixing of the $\uparrow$ and $\downarrow$ branches causes instead
sudden changes of both normal and antinormal angles along each
dispersion. This is the case of both the $0$-\emph{diffusive} spectrum
shown in panel (b) of Fig.~\ref{fig:eigen} (corresponding to panel 5
of Fig. \ref{fig:spect}) as well as the $\omega$-\emph{diffusive}
spectrum shown in panel (c) (and corresponding to the spectrum 7 of
Fig. \ref{fig:spect}), even if the mixing happens in different ways.
In particular, for the $0$-\emph{diffusive} spectrum, different
  branches touch each other in two separated regions in
  $\vect{k}$-space. The branches $u_\downarrow$ and $v_\downarrow$ mix
  together for $k \lesssim 1.5\sqrt{m\mathcal{E}}$, where the angles
  $\theta_{\downarrow, \vect{k}}$ and $\eta_{\downarrow, \vect{k}}$
  coincide. Inside this region, there is a another region where
  also the $u_\uparrow$ and $v_\uparrow$ mix together. Here, also the
  angles $\theta_{\uparrow, \vect{k}}$ and $\eta_{\uparrow, \vect{k}}$
  will coincide. In between these mixing regions, the values of the
  angles undergo a sudden change in value from almost a purely
  left-circularly polarised degree to a right-circularly polarised
  degree.

The last case we analyse is the $\omega$-\emph{diffusive} spectrum
shown in panel (c) of Fig. \ref{fig:eigen}. The difference with the
case previously considered of a $0$-\emph{diffusive} spectrum, lies in
the fact that now all four real part branches do not become degenerate
in the same momentum region and that the degeneracy of $\uparrow$ and
$\downarrow$ branches is allowed at finite energy.
Aside these differences in how $\uparrow$ and $\downarrow$
  branches do mix with each other, we observe in panel (f) a similar
  sudden flip of both normal and anti-normal polar angles along the
  dispersion that we also observed for the $0$-\emph{diffusive}
  spectrum in panel (e).

\begin{figure}
\centering \includegraphics[width=\columnwidth]{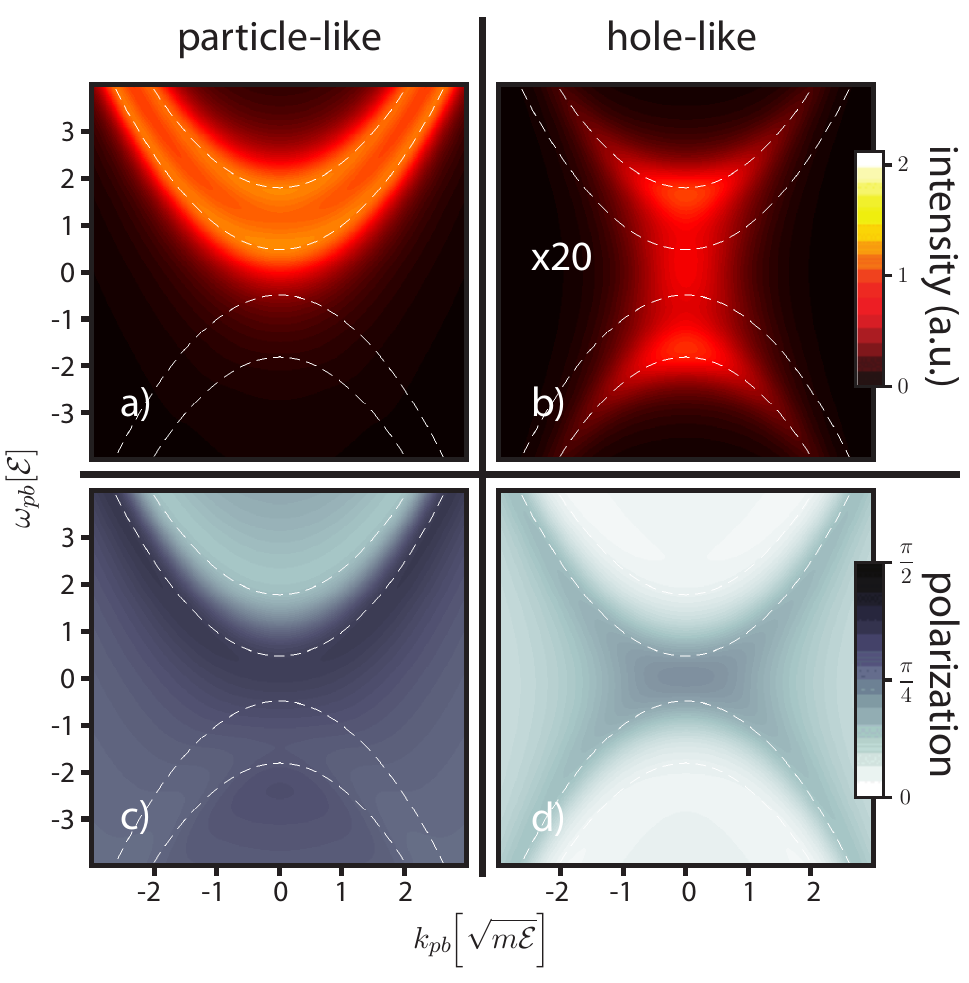}
\caption{(Color online) Response to a probe beam for a \emph{gapped}
  spectrum of excitations. Two-dimensional maps of the particle-like
  $I_{u}$ ( panel (a)) and hole-like $I_{u}$ (panel (b)) intensities,
  and of the polar angles along the $\vect{k}_{pb}$ direction
  $\theta_{u}$ (panel (c)) and the $-\vect{k}_{pb}$ direction
  $\theta_{v}$ (panel (d)) as a function of both the probe momentum
  $\vect{k}_{pb}$ and energy $\omega_{pb}$. The system parameters are
  the same ones fixed in panels (a) and (d) of Fig.~\ref{fig:eigen},
  the probe is linearly polarized ($\theta_{pb}=\pi/4$) and the
  polariton decay rate is set to $\gamma=1.5\mathcal{E}$. White dashed
  lines are the real part of the excitation spectrum $\Re
  \omega^{(a)}_{\vect{k}}$. Note that the intensity of the hole-like
  signal has been multiplied by a factor $20$ with respect to the
  particle-like to obtain a clearer contrast.}
\label{fig:probe_gap}
\end{figure}
\begin{figure}
\centering
\includegraphics[width=\columnwidth]{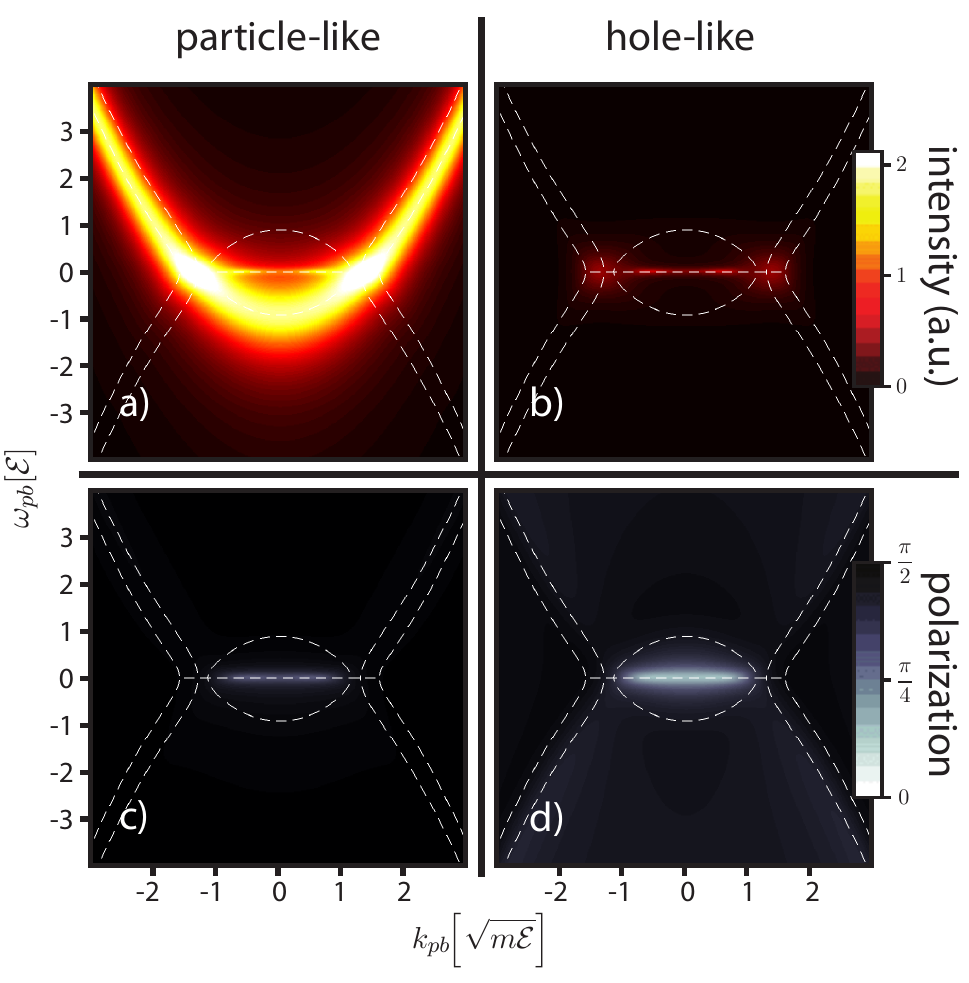}
\caption{Response to a probe beam for a $0$-\emph{diffusive} spectrum
  of excitations. Two-dimensional maps of the particle-like $I_{u}$
  (panel (a)) and hole-like $I_{u}$ (panel (b)) intensities, and of
  the polar angles along the $\vect{k}_{pb}$ direction $\theta_{u}$
  (panel (c)) and the $-\vect{k}_{pb}$ direction $\theta_{v}$ (panel
  (d)) as a function of both the probe momentum $\vect{k}_{pb}$ and
  energy $\omega_{pb}$. The system parameters are the same ones fixed
  in panels (b) and (e) of Fig.~\ref{fig:eigen}, the probe purely
  circularly right-polarized ($\theta_{pb}=\pi/2$) and the polariton
  decay rate is set to $\gamma=1.5\mathcal{E}$. White dashed lines are
  the real part of the excitation spectrum $\Re
  \omega^{(a)}_{\vect{k}}$.}
\label{fig:probe_dif}
\end{figure}
\begin{figure}
\centering \includegraphics[width=\columnwidth]{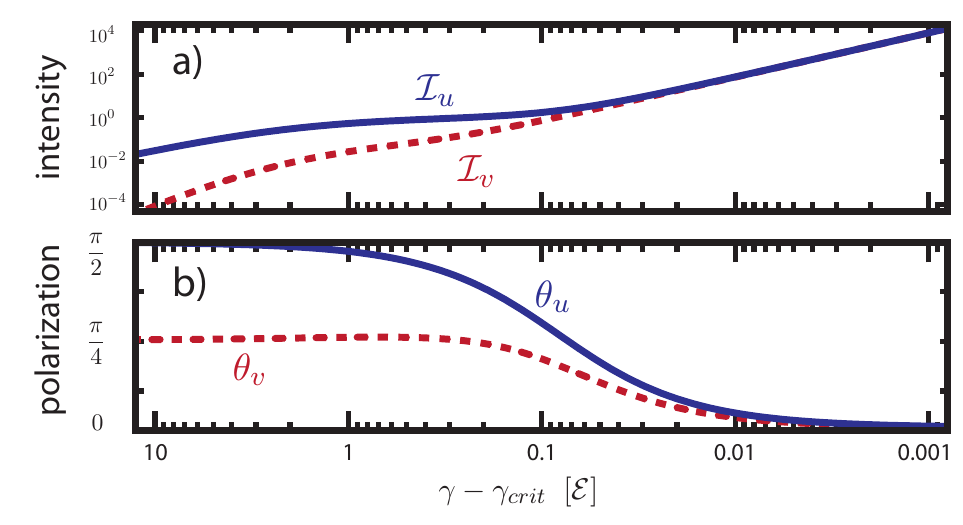}
\caption{Intensities $I_{u,v}$ (panel (a)) and polar angles
  $\theta_{u,v}$ (panel (b)) for the $0$-\emph{diffusive} spectrum of
  excitations shown in Fig.~\ref{fig:probe_dif} at fixed values of the
  probe momentum $\vect{k}_{pb}=0.5\sqrt{m\mathcal{E}}$ and energy
  $\omega_{pb}=0$ as a function of the rescaled polariton decay rate
  $\gamma - \gamma_{crit}$, where $\gamma_{crit} \simeq 1.39\mathcal{E}$.}
\label{fig:gamma}
\end{figure}
%
%
\subsection{Probe response}
\label{sec:probe}
After having discussed the intrinsic properties of the collective
spectrum, including the emission degree of polarisation along the
dispersion of each branch, we derive now the response of the spinor
fluid to an additional probe beam and how this is related to the
intrinsic spectral properties discussed so far.
To this end, we go back to the system linear response~\eqref{eq:invre}
to a weak external probe shined at a direction
$\vect{k}_p+\vect{k}_{pb}$ and an energy $\omega_p + \omega_{pb}$.
Eq.~\eqref{eq:invre} can be easily inverted to give the system
response $\vect{w} = (u_+^{}, v_+^{}, u_-^{}, v_-^{})^T$ in terms of
the probe vector $\vect{f}^{pb} = (f_{+}^{pb}, 0, f_{-}^{pb}, 0)^T$:
\begin{equation}
  \vect{w} = \left(\omega_{pb} \hat{\mathbb{I}} - 
    \hat{\mathcal{L}}_{\vect{k}_{pb}}\right)^{-1}
  \vect{f}^{pb} \; .
\label{eq:respo}
\end{equation}
As explained previously, and also illustrated in the schematic set-up
of Fig.~\ref{fig:schem}, the intensity of the response in the
direction $\vect{k}_p + \vect{k}_{pb}$ will be given by the
particle-like component $|u_{\pm}^{}|^2$, while the one at the
direction $\vect{k}_p - \vect{k}_{pb}$, by the hole-like component
$|v_{\pm}^{}|^2$ --- from now onwards, we will consider the particular
case of a pump shined orthogonally to the cavity plane,
$\vect{k}_p=0$.
From Eq.~\eqref{eq:respo}, we expect that the response will be
enhanced when $(\omega_{pb},\vect{k}_{pb})$ is scanned close to one of
the branches of the collective excitation spectrum, i.e., the
eigenvalues of the Bogoliubov matrix
$\hat{\mathcal{L}}_{\vect{k}_{pb}}$. Because the spectrum is complex,
we expect a broadened enhanced emission --- typically of the order of
the polariton linewidth $\gamma$, with variations in the diffusive
regions in momentum space where different branches touch each other
and the imaginary part of the spectrum deviates from the
polariton lifetime $\gamma$.
While the resolution in energy is limited by the imaginary part of the
spectrum, the resolution in momentum space can only suffer
experimental limitations, such as the angular resolution of the
detection device and the intrinsic angular width of the probe
beam. The probe beam resolution in momentum can be improved by
considering a large enough homogeneous profile~\cite{michiel_OPO}.

We express the probe vector $\vect{f}^{pb}$ in terms of the polar
$\theta_{pb}$ and azimuthal $\phi_{pb}$ angles, quantifying the
polarisation degree of the probe:
\begin{equation}
  \vect{f}^{pb} = |f_{}^{pb}| ( \cos 2\theta_{pb}, 0, e^{i \phi_{pb}}
  \sin 2\theta_{pb}, 0)^T\; ,
\label{eq:prbea}
\end{equation}
where $|f_{}^{pb}|^2$ is the probe beam intensity. We set $\phi_{pb} = 0$.
Similarly, the response $\vect{w}$ can be conveniently parametrised in
terms of the particle-like $I_{u} = |u_+^{}|^2 + |u_-^{}|^2$ and
hole-like $I_{v} = |v_+^{}|^2 + |v_-^{}|^2$ intensities, as well as
the polar angles $\theta_{u, v}$ along the two directions $\pm
\vect{k}_{pb}$:
\begin{align}
  I_u \cos (2 \theta_u) &= |u_+^{}|^2 - |u_-^{}|^2\\
  I_v \cos (2 \theta_v) &= |v_+^{}|^2 - |v_-^{}|^2\; .
\end{align}
Finally, as already discussed, we assume that the pump induces a
mean-field state with $\phi_0 = 0$ and $\theta_0 \in [0,
  \pi/4]$~\eqref{eq:mfpol}.

We plot in Figs.~\ref{fig:probe_gap} and~\ref{fig:probe_dif} the
  two-dimensional maps for both intensities $I_{u,v}$ and polar angles
  $\theta_{u,v}$ for the response to a probe beam by scanning
  different values of the the probe energy $\omega_{pb}$ and momentum
  $\vect{k}_{pb}$. We choose the particular case of a \emph{gapped}
  spectrum (Fig.~\ref{fig:probe_gap}), corresponding to the same
  conditions as panels (a) and (d) of Fig.~\ref{fig:eigen} and the
  case of a $0$-\emph{diffusive} spectrum (Fig.~\ref{fig:probe_dif}),
  corresponding to the same conditions as panels (b) and (e) of
  Fig.~\ref{fig:eigen}.
We first note that, as expected, the probe beam~\eqref{eq:prbea}
  has only a finite strength in the particle-like channels and cannot
  directly excite hole-like quasi-particles. Nevertheless, the
  response has a finite emission intensity also for the hole-like
  branches because of the interactions mixing together particle and
  hole degrees of freedom. As previously illustrated in
  Eq.~\eqref{eq:freed} particle and hole modes asymptotically decouple
  one from the other at large momenta. We thus expect that the
  response emission intensity quickly drops to zero for all hole-like
  branches. In panel (b) of Fig.~\ref{fig:probe_gap} this behaviour is
  clearly visible, as the emission intensity of the hole branches is
  strongly reduced with respect to the one of the particle branches.

In the previous section we have seen that each branch of the
collective excitation spectrum is characterised by an intrinsic
degrees of polarization, quantified by the normal
($\theta_{\uparrow,\downarrow}$) and anti-normal
($\eta_{\uparrow,\downarrow}$) polar angles. 
In Fig.~\ref{fig:probe_gap}, the \emph{gapped} spectrum is probed with
a linearly polarized beam, $\theta_{pb}=\pi/4$. Here, we obtained an
enhanced emission when the probe is in resonance with either the
$u_\uparrow$ or the $u_\downarrow$ branch. We remind that for a
\emph{gapped} spectrum there is no mixing between the $\uparrow$ and
$\downarrow$ degrees of freedom, and thus the
$\theta_{\uparrow,\downarrow}$ angles only weakly deviate from their
asymptotic values at large $\vect{k}$ (see Fig.~\ref{fig:eigen}), in
this particular case almost purely circularly left or right
polarized. Hence, one expects that the emission intensity for the
$u_\uparrow$ and $u_\downarrow$ branches is comparable, since the
linearly polarised probe couples identically to left- and
right-polarized modes.
In panel (c) of Fig.~\ref{fig:probe_gap} we observe the polarization
to undergo a rotation when in resonance with one of the spectral
branches. As expected, probing at resonance with the $u_\uparrow$
branch results in a largely left-polarized emission, while the
$u_\downarrow$ branch induces a right-circular polarization.

Interesting effects are observed for the case of diffusive spectra,
and we show in particular the case of a $0$-\emph{diffusive} spectrum
in Fig.~\ref{fig:probe_dif}. For this case we consider the case of a
right-circularly polarized probe beam ($\theta_{pb}=\pi/2$) so that we
have a $u_{\downarrow}$ branch emitting much stronger than the
$u_{\uparrow}$ branch (see panel (a)
of~Fig. \ref{fig:probe_dif}). However, around $\vect{k}_p=0$, where
the spectral branches mixed one the other in a diffusive region, we
see that the resonant branch is not $u_{\downarrow}$, rather
$v_{\uparrow}$.
Here, the strongest emission intensity comes from the $u_{\downarrow}$
branch at large momenta and from the $v_\uparrow$ branch inside the
diffusive region at small $\vect{k}$.
Because of the mixing between $\uparrow$ and $\downarrow$ degrees of
freedom, we have observed a sudden spin flip for the degree of
polarisation of the eigenvectors around $|\vect{k}|\sim
1.2\sqrt{m\mathcal{E}}$. This is the reason for the resonant emission
transfers from $u_{\downarrow}$ to $v_{\uparrow}$ at small
$\vect{k}$.
From panel (b) of Fig.~\ref{fig:probe_dif} it is clear that, in the
beam transferred in the $-\vect{k}_{pb}$ direction, the emission is
the strongest in the diffusive momentum ring at $\omega_p=0$ where
$u_\downarrow$ and $ v_\downarrow$ are mixed. 
The stronger intensity for these modes can be explained in terms of an
onset of parametric amplification. In fact, in this diffusive region
the imaginary part of the spectrum part deviates from the value of the
constant decay rate and there is an additional contribution from the
negative argument of the outer root in Eq.~\eqref{eq:collsp} --- see
panel 5 in Fig.~\ref{fig:spect}. For this reason, these modes acquire
a longer lifetime, and therefore undergo enhanced
scattering. Similarly, in panel (a) of Fig.~\ref{fig:probe_dif}, one
can appreciate an increase in transmission in the same diffusive
region.
In addition, there is also an increased emission on the diffusive
region at $\omega_p=0$ where the $u_\downarrow$ and $v_\downarrow$
branches are degenerate. Although, for these values of the momentum,
these branches are circularly left polarized, the inter-spin
interaction coupling the $\uparrow$ and $\downarrow$ modes, combined
with the strong parametric amplification, still leads to a strong
emission.
It is interesting to note that in previous work~\cite{neg_drag}, the
proximity to an instability to a parametric scattering regime has been
related to the occurrence of a negative drag force when a resonantly
pumped polariton fluid scatters against a localised defect. Here, the
same phenomenon leads to enhanced emission of parametrically amplified
modes when the system is probed with an additional weak laser.

In panels (c) [(d)] of Fig~\ref{fig:probe_dif} we plot the azimuthal
polarization angle for particle-like (hole-like)transmitted signal at
$\vect{k}_{pb}$ ($-\vect{k}_{pb}$). The resonant branch is
$u_\downarrow$ at large $\vect{k}$ and $v_\uparrow$ for small values
of the momentum and strongly emits circularly right polarized light
--- $\theta^u_{pb}\simeq \theta^v_{pb}\simeq \pi/2$.
Yet the parametrically amplified region around $\vect{k}_p=0$ emits
both in the transmitted $\vect{k}_{pb}$ and $-\vect{k}_{pb}$
directions an almost purely circularly left polarized light. Hence,
the parametric amplification of the mixed $u_\downarrow$ and
$v_\downarrow$ mode causes the left-polarized incoming probe light to
undergo a spin flip when interacting with the polariton sample.  
We show in Fig.~\ref{fig:gamma} that $\theta_{pb}^{u,v}$ tend to $0$
(i.e., almost pure left-polarization), when the parametric amplified
mode $\vect{k}_p=0.5\sqrt{m\mathcal{E}}$, probed with $\omega_{pb}=0$,
is brought closer to resonance by varying the polariton decay time
$\gamma \to \gamma_{crit}^{-}$ --- here $\gamma_{crit}$ is the minimum
value of the polariton decay rate required for the system stability,
i.e., $\Im \omega<0$.
The emission intensity (panel (a) of Fig.~\ref{fig:gamma}) diverges as
$1/(\gamma-\gamma_{crit})^2$ for both particle-like and hole-like
signals when they are brought close to resonance.
This can be understood from the expression~\eqref{eq:respo}, where we
see that the right-hand side becomes singular if $\omega_{pb}$ equals
$\Re\omega^{(a)}_{\vect{k}_{pb}}$ and at the same time $\Im
\omega^{(a)}_{\vect{k}_{pb}} \to 0^{-}$.
In this limit, the response $\vect{w}$ coincides with the eigenvector
$\vect{x}^{(a)}$ of the Bogoliubov matrix
$\hat{\mathcal{L}}_{\vect{k}_{pb}}$. Here, this is the case for
$\vect{x}^{u_-}$, so that the polarization $\theta_{pb}$ of the
response approaches
$\theta_{\downarrow}|_{\vect{k}=0}=\eta_{\downarrow}|_{\vect{k}=0}\simeq
0.08$, as it can be seen in panel (b) of Fig.~\ref{fig:gamma}. Note
that $\theta_\downarrow=\eta_{\downarrow}$ exactly in the diffusive
disk around $\vect{k}=0$, where $u_\downarrow$ and $v_\downarrow$
coincide. In Fig.~\ref{fig:gamma}, we have measured the decay rate
$\gamma$ in units of self-interaction energy
$\mathcal{E}$. 

Experimentally, however, $\gamma$ is fixed for a given
microcavity, but the ratio $\gamma/\mathcal{E}$ can be tuned by
varying the laser pump power, $|f^p|^2$ in Eq.~\eqref{eq:pumpo}. One
should then calculate the mean-field equations~\eqref{eq:meanf} to
derive the polariton spin densities $|\psi_{\pm}|^2$ and thus the
self-interaction energy $\mathcal{E}$.
By varying the tilting angle of the probe beam with respect to the pump, one can scan through $\vect{k}$-space. As depicted on Fig. \ref{fig:schem}, also the detectors should be placed accordingly: one at the same angle as the probe and the other at the mirrored angle, to detect the particle and the hole-like signal respectively. At each position of the probe, a vertical slice on the response figures can be reconstructed by changing the probe frequency $\omega_{pb}$ and measuring the intensity and polarization at both detectors. 

\section{Conclusions}
\label{sec:concl}
We have analytically derived the spectrum of elementary excitations
for a spinor polariton fluid in the linear response approximation
scheme. For fixed interaction stength, the spectra can be classified
in terms of two dimensionless parameters only: mean-field polarization
angle and the renormalised pump detuning. Even though there is a large
variety of possible spectra, we identify three major classes,
\emph{gapped}, $0$-\emph{diffusive}, and $\omega$-\emph{diffusive},
depending how the opposite polarisation spectral branches mix together
and at which energy. For $0$-\emph{diffusive} the mixing happens at
zero energies, for $\omega$-\emph{diffusive} it happens at finite
energy. Interestingly, only the mean-field polarization is sufficient
to distinguish between these two different diffusive-like spectra. We
show that the mixing of $\uparrow$ and $\downarrow$ branches
characterises sudden spin flips of the intristic degree of
polarisation along the branches for both diffusive-like spectra.
We have characterised the response of the system to an external probe
in terms of the spectral intrinsic properties. In particular, we have
shown that the intrinsic polarization of an elementary excitation is
reflected in the trasmitted signal of a probe beam experiment. For
\emph{gapped} spectra the degree of polarisation varies only very
weakly along each branches. In contrast, for both $0$-\emph{diffusive}
and $\omega$-\emph{diffusive} spectra, the strong mixing between
opposite polarisation branches at small momenta leads to a spin flip
of the transmitted degree of polarisation along the branch. The closer
the polariton spinor fluid is to a parametric instability, the larger
the amount of spin flip is, independently of the degree of polarisation
of the probing beam.

Recently numerous fascinating results have been achieved in the study
of the response of a spinor polariton fluid to a magnetic field, such
as the spin Meissner effect\cite{spin_meissner} and effective magnetic
monopoles\cite{soln_monopoles}. As a future perspective, it could be
interesting to include the Zeeman-splitting terms in our model and
study their influence on the spectrum of excitations. In addition,
also effects of disorder
\cite{phase_disorder,suppression_superfluidity} and TE-TM splitting
could be incorporated.
%

\acknowledgments 
MVR gratefully acknowledges support in the form of a Ph. D. fellowship
of the Research Foundation - Flanders (FWO). MW acknowledges financial
support from the FWO-Odysseus program.  F.M.M. acknowledges financial
support from the Ministerio de Econom\'ia y Competitividad (MINECO)
(Contract No. MAT2011-22997) and the Comunidad Autonoma de Madrid
(CAM) (Contract No. S-2009/ESP-1503).
%


\begin{thebibliography}{31}%
\makeatletter
\providecommand \@ifxundefined [1]{%
 \@ifx{#1\undefined}
}%
\providecommand \@ifnum [1]{%
 \ifnum #1\expandafter \@firstoftwo
 \else \expandafter \@secondoftwo
 \fi
}%
\providecommand \@ifx [1]{%
 \ifx #1\expandafter \@firstoftwo
 \else \expandafter \@secondoftwo
 \fi
}%
\providecommand \natexlab [1]{#1}%
\providecommand \enquote  [1]{``#1''}%
\providecommand \bibnamefont  [1]{#1}%
\providecommand \bibfnamefont [1]{#1}%
\providecommand \citenamefont [1]{#1}%
\providecommand \href@noop [0]{\@secondoftwo}%
\providecommand \href [0]{\begingroup \@sanitize@url \@href}%
\providecommand \@href[1]{\@@startlink{#1}\@@href}%
\providecommand \@@href[1]{\endgroup#1\@@endlink}%
\providecommand \@sanitize@url [0]{\catcode `\\12\catcode `\$12\catcode
  `\&12\catcode `\#12\catcode `\^12\catcode `\_12\catcode `\%12\relax}%
\providecommand \@@startlink[1]{}%
\providecommand \@@endlink[0]{}%
\providecommand \url  [0]{\begingroup\@sanitize@url \@url }%
\providecommand \@url [1]{\endgroup\@href {#1}{\urlprefix }}%
\providecommand \urlprefix  [0]{URL }%
\providecommand \Eprint [0]{\href }%
\providecommand \doibase [0]{http://dx.doi.org/}%
\providecommand \selectlanguage [0]{\@gobble}%
\providecommand \bibinfo  [0]{\@secondoftwo}%
\providecommand \bibfield  [0]{\@secondoftwo}%
\providecommand \translation [1]{[#1]}%
\providecommand \BibitemOpen [0]{}%
\providecommand \bibitemStop [0]{}%
\providecommand \bibitemNoStop [0]{.\EOS\space}%
\providecommand \EOS [0]{\spacefactor3000\relax}%
\providecommand \BibitemShut  [1]{\csname bibitem#1\endcsname}%
\let\auto@bib@innerbib\@empty
\bibitem [{\citenamefont {Kavokin}\ \emph {et~al.}(2007)\citenamefont
  {Kavokin}, \citenamefont {Baumberg}, \citenamefont {Malpuech},\ and\
  \citenamefont {Laussy}}]{kavokin_laussy}%
  \BibitemOpen
  \bibfield  {author} {\bibinfo {author} {\bibfnamefont {A.~V.}\ \bibnamefont
  {Kavokin}}, \bibinfo {author} {\bibfnamefont {J.~J.}\ \bibnamefont
  {Baumberg}}, \bibinfo {author} {\bibfnamefont {G.}~\bibnamefont {Malpuech}},
  \ and\ \bibinfo {author} {\bibfnamefont {F.~P.}\ \bibnamefont {Laussy}},\
  }\href@noop {} {\emph {\bibinfo {title} {{Microcavities}}}}\ (\bibinfo
  {publisher} {Oxford University Press},\ \bibinfo {address} {Oxford},\
  \bibinfo {year} {2007})\BibitemShut {NoStop}%
\bibitem [{\citenamefont {Keeling}\ \emph {et~al.}(2006)\citenamefont
  {Keeling}, \citenamefont {Marchetti}, \citenamefont {Szyma\'{n}ska},\ and\
  \citenamefont {Littlewood}}]{keeling_review07}%
  \BibitemOpen
  \bibfield  {author} {\bibinfo {author} {\bibfnamefont {J.}~\bibnamefont
  {Keeling}}, \bibinfo {author} {\bibfnamefont {F.~M.}\ \bibnamefont
  {Marchetti}}, \bibinfo {author} {\bibfnamefont {M.~H.}\ \bibnamefont
  {Szyma\'{n}ska}}, \ and\ \bibinfo {author} {\bibfnamefont {P.~B.}\
  \bibnamefont {Littlewood}},\ }\href@noop {} {\bibfield  {journal} {\bibinfo
  {journal} {Semicond. Sci. Technol.}\ }\textbf {\bibinfo {volume} {22}},\
  \bibinfo {pages} {R1} (\bibinfo {year} {2006})}\BibitemShut {NoStop}%
\bibitem [{\citenamefont {Keeling}\ and\ \citenamefont
  {Berloff}(2011)}]{keeling_berloff_review}%
  \BibitemOpen
  \bibfield  {author} {\bibinfo {author} {\bibfnamefont {J.}~\bibnamefont
  {Keeling}}\ and\ \bibinfo {author} {\bibfnamefont {N.~G.}\ \bibnamefont
  {Berloff}},\ }\href {\doibase 10.1080/00107514.2010.550120} {\bibfield
  {journal} {\bibinfo  {journal} {Contemporary Physics}\ }\textbf {\bibinfo
  {volume} {52}},\ \bibinfo {pages} {131} (\bibinfo {year} {2011})},\ \Eprint
  {http://arxiv.org/abs/http://dx.doi.org/10.1080/00107514.2010.550120}
  {http://dx.doi.org/10.1080/00107514.2010.550120} \BibitemShut {NoStop}%
\bibitem [{\citenamefont {Carusotto}\ and\ \citenamefont
  {Ciuti}(2013)}]{iac_review}%
  \BibitemOpen
  \bibfield  {author} {\bibinfo {author} {\bibfnamefont {I.}~\bibnamefont
  {Carusotto}}\ and\ \bibinfo {author} {\bibfnamefont {C.}~\bibnamefont
  {Ciuti}},\ }\href {\doibase 10.1103/RevModPhys.85.299} {\bibfield  {journal}
  {\bibinfo  {journal} {Rev. Mod. Phys.}\ }\textbf {\bibinfo {volume} {85}},\
  \bibinfo {pages} {299} (\bibinfo {year} {2013})}\BibitemShut {NoStop}%
\bibitem [{\citenamefont {Carusotto}\ and\ \citenamefont
  {Ciuti}(2004)}]{carusotto04}%
  \BibitemOpen
  \bibfield  {author} {\bibinfo {author} {\bibfnamefont {I.}~\bibnamefont
  {Carusotto}}\ and\ \bibinfo {author} {\bibfnamefont {C.}~\bibnamefont
  {Ciuti}},\ }\href@noop {} {\bibfield  {journal} {\bibinfo  {journal} {Phys.
  Rev. Lett.}\ }\textbf {\bibinfo {volume} {93}},\ \bibinfo {eid} {166401}
  (\bibinfo {year} {2004})}\BibitemShut {NoStop}%
\bibitem [{\citenamefont {Ciuti}\ and\ \citenamefont
  {Carusotto}(2005)}]{ciuti_05}%
  \BibitemOpen
  \bibfield  {author} {\bibinfo {author} {\bibfnamefont {C.}~\bibnamefont
  {Ciuti}}\ and\ \bibinfo {author} {\bibfnamefont {I.}~\bibnamefont
  {Carusotto}},\ }\href@noop {} {\bibfield  {journal} {\bibinfo  {journal}
  {physica status solidi (b)}\ }\textbf {\bibinfo {volume} {242}},\ \bibinfo
  {pages} {2224} (\bibinfo {year} {2005})}\BibitemShut {NoStop}%
\bibitem [{\citenamefont {Amo}\ \emph {et~al.}(2009)\citenamefont {Amo},
  \citenamefont {Lefr\`ere}, \citenamefont {Pigeon}, \citenamefont {Adrados},
  \citenamefont {Ciuti}, \citenamefont {Carusotto}, \citenamefont {Houdr\'e},
  \citenamefont {Giacobino},\ and\ \citenamefont {Bramati}}]{amo09_b}%
  \BibitemOpen
  \bibfield  {author} {\bibinfo {author} {\bibfnamefont {A.}~\bibnamefont
  {Amo}}, \bibinfo {author} {\bibfnamefont {J.}~\bibnamefont {Lefr\`ere}},
  \bibinfo {author} {\bibfnamefont {S.}~\bibnamefont {Pigeon}}, \bibinfo
  {author} {\bibfnamefont {C.}~\bibnamefont {Adrados}}, \bibinfo {author}
  {\bibfnamefont {C.}~\bibnamefont {Ciuti}}, \bibinfo {author} {\bibfnamefont
  {I.}~\bibnamefont {Carusotto}}, \bibinfo {author} {\bibfnamefont
  {R.}~\bibnamefont {Houdr\'e}}, \bibinfo {author} {\bibfnamefont
  {E.}~\bibnamefont {Giacobino}}, \ and\ \bibinfo {author} {\bibfnamefont
  {A.}~\bibnamefont {Bramati}},\ }\href@noop {} {\bibfield  {journal} {\bibinfo
   {journal} {Nat. Phys.}\ }\textbf {\bibinfo {volume} {5}},\ \bibinfo {pages}
  {805} (\bibinfo {year} {2009})}\BibitemShut {NoStop}%
\bibitem [{\citenamefont {Pigeon}\ \emph {et~al.}(2011)\citenamefont {Pigeon},
  \citenamefont {Carusotto},\ and\ \citenamefont {Ciuti}}]{pigeon_11}%
  \BibitemOpen
  \bibfield  {author} {\bibinfo {author} {\bibfnamefont {S.}~\bibnamefont
  {Pigeon}}, \bibinfo {author} {\bibfnamefont {I.}~\bibnamefont {Carusotto}}, \
  and\ \bibinfo {author} {\bibfnamefont {C.}~\bibnamefont {Ciuti}},\
  }\href@noop {} {\bibfield  {journal} {\bibinfo  {journal} {Phys. Rev. B}\
  }\textbf {\bibinfo {volume} {83}},\ \bibinfo {pages} {144513} (\bibinfo
  {year} {2011})}\BibitemShut {NoStop}%
\bibitem [{\citenamefont {Amo}\ \emph {et~al.}(2011)\citenamefont {Amo},
  \citenamefont {Pigeon}, \citenamefont {Sanvitto}, \citenamefont {Sala},
  \citenamefont {Hivet}, \citenamefont {Carusotto}, \citenamefont {Pisanello},
  \citenamefont {Lem\'enager}, \citenamefont {Houdr\'e}, \citenamefont
  {Giacobino}, \citenamefont {Ciuti},\ and\ \citenamefont
  {Bramati}}]{amo_science11}%
  \BibitemOpen
  \bibfield  {author} {\bibinfo {author} {\bibfnamefont {A.}~\bibnamefont
  {Amo}}, \bibinfo {author} {\bibfnamefont {S.}~\bibnamefont {Pigeon}},
  \bibinfo {author} {\bibfnamefont {D.}~\bibnamefont {Sanvitto}}, \bibinfo
  {author} {\bibfnamefont {V.~G.}\ \bibnamefont {Sala}}, \bibinfo {author}
  {\bibfnamefont {R.}~\bibnamefont {Hivet}}, \bibinfo {author} {\bibfnamefont
  {I.}~\bibnamefont {Carusotto}}, \bibinfo {author} {\bibfnamefont
  {F.}~\bibnamefont {Pisanello}}, \bibinfo {author} {\bibfnamefont
  {G.}~\bibnamefont {Lem\'enager}}, \bibinfo {author} {\bibfnamefont
  {R.}~\bibnamefont {Houdr\'e}}, \bibinfo {author} {\bibfnamefont
  {E.}~\bibnamefont {Giacobino}}, \bibinfo {author} {\bibfnamefont
  {C.}~\bibnamefont {Ciuti}}, \ and\ \bibinfo {author} {\bibfnamefont
  {A.}~\bibnamefont {Bramati}},\ }\href@noop {} {\bibfield  {journal} {\bibinfo
   {journal} {Science}\ }\textbf {\bibinfo {volume} {332}},\ \bibinfo {pages}
  {1167} (\bibinfo {year} {2011})}\BibitemShut {NoStop}%
\bibitem [{\citenamefont {Nardin}\ \emph {et~al.}(2011)\citenamefont {Nardin},
  \citenamefont {Grosso}, \citenamefont {Leger}, \citenamefont {Pietka},
  \citenamefont {Morier-Genoud},\ and\ \citenamefont
  {Deveaud-Pl\'edran}}]{nardin11}%
  \BibitemOpen
  \bibfield  {author} {\bibinfo {author} {\bibfnamefont {G.}~\bibnamefont
  {Nardin}}, \bibinfo {author} {\bibfnamefont {G.}~\bibnamefont {Grosso}},
  \bibinfo {author} {\bibfnamefont {Y.}~\bibnamefont {Leger}}, \bibinfo
  {author} {\bibfnamefont {B.}~\bibnamefont {Pietka}}, \bibinfo {author}
  {\bibfnamefont {F.}~\bibnamefont {Morier-Genoud}}, \ and\ \bibinfo {author}
  {\bibfnamefont {B.}~\bibnamefont {Deveaud-Pl\'edran}},\ }\href {\doibase
  doi:10.1038/nphys1959} {\bibfield  {journal} {\bibinfo  {journal} {Nature
  Physics}\ }\textbf {\bibinfo {volume} {7}},\ \bibinfo {pages} {635} (\bibinfo
  {year} {2011})}\BibitemShut {NoStop}%
\bibitem [{\citenamefont {Sanvitto}\ \emph {et~al.}(2011)\citenamefont
  {Sanvitto}, \citenamefont {Pigeon}, \citenamefont {Amo}, \citenamefont
  {Ballarini}, \citenamefont {Giorgi}, \citenamefont {Carusotto}, \citenamefont
  {Hivet}, \citenamefont {Pisanello}, \citenamefont {Sala}, \citenamefont
  {Soares-Guimaraes}, \citenamefont {Houdre}, \citenamefont {Giacobino},
  \citenamefont {Ciuti}, \citenamefont {Bramati},\ and\ \citenamefont
  {Gigli}}]{sanvitto11}%
  \BibitemOpen
  \bibfield  {author} {\bibinfo {author} {\bibfnamefont {D.}~\bibnamefont
  {Sanvitto}}, \bibinfo {author} {\bibfnamefont {S.}~\bibnamefont {Pigeon}},
  \bibinfo {author} {\bibfnamefont {A.}~\bibnamefont {Amo}}, \bibinfo {author}
  {\bibfnamefont {D.}~\bibnamefont {Ballarini}}, \bibinfo {author}
  {\bibfnamefont {M.~D.}\ \bibnamefont {Giorgi}}, \bibinfo {author}
  {\bibfnamefont {I.}~\bibnamefont {Carusotto}}, \bibinfo {author}
  {\bibfnamefont {R.}~\bibnamefont {Hivet}}, \bibinfo {author} {\bibfnamefont
  {F.}~\bibnamefont {Pisanello}}, \bibinfo {author} {\bibfnamefont {V.~G.}\
  \bibnamefont {Sala}}, \bibinfo {author} {\bibfnamefont {P.~S.}\ \bibnamefont
  {Soares-Guimaraes}}, \bibinfo {author} {\bibfnamefont {R.}~\bibnamefont
  {Houdre}}, \bibinfo {author} {\bibfnamefont {E.}~\bibnamefont {Giacobino}},
  \bibinfo {author} {\bibfnamefont {C.}~\bibnamefont {Ciuti}}, \bibinfo
  {author} {\bibfnamefont {A.}~\bibnamefont {Bramati}}, \ and\ \bibinfo
  {author} {\bibfnamefont {G.}~\bibnamefont {Gigli}},\ }\href@noop {}
  {\bibfield  {journal} {\bibinfo  {journal} {Nature Photonics}\ }\textbf
  {\bibinfo {volume} {5}},\ \bibinfo {pages} {610} (\bibinfo {year}
  {2011})}\BibitemShut {NoStop}%
\bibitem [{\citenamefont {Van~Regemortel}\ and\ \citenamefont
  {Wouters}(2014)}]{neg_drag}%
  \BibitemOpen
  \bibfield  {author} {\bibinfo {author} {\bibfnamefont {M.}~\bibnamefont
  {Van~Regemortel}}\ and\ \bibinfo {author} {\bibfnamefont {M.}~\bibnamefont
  {Wouters}},\ }\href {\doibase 10.1103/PhysRevB.89.085303} {\bibfield
  {journal} {\bibinfo  {journal} {Phys. Rev. B}\ }\textbf {\bibinfo {volume}
  {89}},\ \bibinfo {pages} {085303} (\bibinfo {year} {2014})}\BibitemShut
  {NoStop}%
\bibitem [{\citenamefont {Gippius}\ \emph {et~al.}(2007)\citenamefont
  {Gippius}, \citenamefont {Shelykh}, \citenamefont {Solnyshkov}, \citenamefont
  {Gavrilov}, \citenamefont {Rubo}, \citenamefont {Kavokin}, \citenamefont
  {Tikhodeev},\ and\ \citenamefont {Malpuech}}]{gippius_multistability}%
  \BibitemOpen
  \bibfield  {author} {\bibinfo {author} {\bibfnamefont {N.}~\bibnamefont
  {Gippius}}, \bibinfo {author} {\bibfnamefont {I.}~\bibnamefont {Shelykh}},
  \bibinfo {author} {\bibfnamefont {D.}~\bibnamefont {Solnyshkov}}, \bibinfo
  {author} {\bibfnamefont {S.}~\bibnamefont {Gavrilov}}, \bibinfo {author}
  {\bibfnamefont {Y.}~\bibnamefont {Rubo}}, \bibinfo {author} {\bibfnamefont
  {A.}~\bibnamefont {Kavokin}}, \bibinfo {author} {\bibfnamefont
  {S.}~\bibnamefont {Tikhodeev}}, \ and\ \bibinfo {author} {\bibfnamefont
  {G.}~\bibnamefont {Malpuech}},\ }\href {\doibase
  10.1103/PhysRevLett.98.236401} {\bibfield  {journal} {\bibinfo  {journal}
  {Phys. Rev. Lett.}\ }\textbf {\bibinfo {volume} {98}},\ \bibinfo {pages}
  {236401} (\bibinfo {year} {2007})}\BibitemShut {NoStop}%
\bibitem [{\citenamefont {Sarkar}\ \emph {et~al.}(2010)\citenamefont {Sarkar},
  \citenamefont {Gavrilov}, \citenamefont {Sich}, \citenamefont {Quilter},
  \citenamefont {Bradley}, \citenamefont {Gippius}, \citenamefont {Guda},
  \citenamefont {Kulakovskii}, \citenamefont {Skolnick},\ and\ \citenamefont
  {Krizhanovskii}}]{sarkar_mult}%
  \BibitemOpen
  \bibfield  {author} {\bibinfo {author} {\bibfnamefont {D.}~\bibnamefont
  {Sarkar}}, \bibinfo {author} {\bibfnamefont {S.~S.}\ \bibnamefont
  {Gavrilov}}, \bibinfo {author} {\bibfnamefont {M.}~\bibnamefont {Sich}},
  \bibinfo {author} {\bibfnamefont {J.~H.}\ \bibnamefont {Quilter}}, \bibinfo
  {author} {\bibfnamefont {R.~A.}\ \bibnamefont {Bradley}}, \bibinfo {author}
  {\bibfnamefont {N.~A.}\ \bibnamefont {Gippius}}, \bibinfo {author}
  {\bibfnamefont {K.}~\bibnamefont {Guda}}, \bibinfo {author} {\bibfnamefont
  {V.~D.}\ \bibnamefont {Kulakovskii}}, \bibinfo {author} {\bibfnamefont
  {M.~S.}\ \bibnamefont {Skolnick}}, \ and\ \bibinfo {author} {\bibfnamefont
  {D.~N.}\ \bibnamefont {Krizhanovskii}},\ }\href {\doibase
  10.1103/PhysRevLett.105.216402} {\bibfield  {journal} {\bibinfo  {journal}
  {Phys. Rev. Lett.}\ }\textbf {\bibinfo {volume} {105}},\ \bibinfo {pages}
  {216402} (\bibinfo {year} {2010})}\BibitemShut {NoStop}%
\bibitem [{\citenamefont {Amo}\ \emph {et~al.}(2010)\citenamefont {Amo},
  \citenamefont {Liew}, \citenamefont {Adrados}, \citenamefont {Houdré},
  \citenamefont {Giacobino}, \citenamefont {Kavokin},\ and\ \citenamefont
  {Bramati}}]{nature_polswitch}%
  \BibitemOpen
  \bibfield  {author} {\bibinfo {author} {\bibfnamefont {A.}~\bibnamefont
  {Amo}}, \bibinfo {author} {\bibfnamefont {T.~C.~H.}\ \bibnamefont {Liew}},
  \bibinfo {author} {\bibfnamefont {C.}~\bibnamefont {Adrados}}, \bibinfo
  {author} {\bibfnamefont {R.}~\bibnamefont {Houdré}}, \bibinfo {author}
  {\bibfnamefont {E.}~\bibnamefont {Giacobino}}, \bibinfo {author}
  {\bibfnamefont {A.~V.}\ \bibnamefont {Kavokin}}, \ and\ \bibinfo {author}
  {\bibfnamefont {A.}~\bibnamefont {Bramati}},\ }\href {\doibase
  10.1038/nphoton.2010.79} {\bibfield  {journal} {\bibinfo  {journal} {Nature
  Photonics}\ }\textbf {\bibinfo {volume} {4}},\ \bibinfo {pages} {361}
  (\bibinfo {year} {2010})}\BibitemShut {NoStop}%
\bibitem [{\citenamefont {Ostatnick\'y}\ \emph {et~al.}(2010)\citenamefont
  {Ostatnick\'y}, \citenamefont {Shelykh},\ and\ \citenamefont
  {Kavokin}}]{kavokin_logic}%
  \BibitemOpen
  \bibfield  {author} {\bibinfo {author} {\bibfnamefont {T.}~\bibnamefont
  {Ostatnick\'y}}, \bibinfo {author} {\bibfnamefont {I.}~\bibnamefont
  {Shelykh}}, \ and\ \bibinfo {author} {\bibfnamefont {A.}~\bibnamefont
  {Kavokin}},\ }\href {\doibase 10.1103/PhysRevB.81.125319} {\bibfield
  {journal} {\bibinfo  {journal} {Phys. Rev. B}\ }\textbf {\bibinfo {volume}
  {81}},\ \bibinfo {pages} {125319} (\bibinfo {year} {2010})}\BibitemShut
  {NoStop}%
\bibitem [{\citenamefont {Takemura}\ \emph {et~al.}(2014)\citenamefont
  {Takemura}, \citenamefont {Trebaol}, \citenamefont {Wouters}, \citenamefont
  {Portella-Oberli},\ and\ \citenamefont {Deveaud}}]{takemura2014}%
  \BibitemOpen
  \bibfield  {author} {\bibinfo {author} {\bibfnamefont {N.}~\bibnamefont
  {Takemura}}, \bibinfo {author} {\bibfnamefont {S.}~\bibnamefont {Trebaol}},
  \bibinfo {author} {\bibfnamefont {M.}~\bibnamefont {Wouters}}, \bibinfo
  {author} {\bibfnamefont {M.~T.}\ \bibnamefont {Portella-Oberli}}, \ and\
  \bibinfo {author} {\bibfnamefont {B.}~\bibnamefont {Deveaud}},\ }\href@noop
  {} {\bibfield  {journal} {\bibinfo  {journal} {Nat Phys}\ }\textbf {\bibinfo
  {volume} {10}},\ \bibinfo {pages} {500} (\bibinfo {year} {2014})}\BibitemShut
  {NoStop}%
\bibitem [{\citenamefont {Solnyshkov}\ \emph {et~al.}(2008)\citenamefont
  {Solnyshkov}, \citenamefont {Shelykh}, \citenamefont {Gippius}, \citenamefont
  {Kavokin},\ and\ \citenamefont {Malpuech}}]{malpuech_disp}%
  \BibitemOpen
  \bibfield  {author} {\bibinfo {author} {\bibfnamefont {D.}~\bibnamefont
  {Solnyshkov}}, \bibinfo {author} {\bibfnamefont {I.}~\bibnamefont {Shelykh}},
  \bibinfo {author} {\bibfnamefont {N.}~\bibnamefont {Gippius}}, \bibinfo
  {author} {\bibfnamefont {A.}~\bibnamefont {Kavokin}}, \ and\ \bibinfo
  {author} {\bibfnamefont {G.}~\bibnamefont {Malpuech}},\ }\href {\doibase
  10.1103/PhysRevB.77.045314} {\bibfield  {journal} {\bibinfo  {journal} {Phys.
  Rev. B}\ }\textbf {\bibinfo {volume} {77}},\ \bibinfo {pages} {045314}
  (\bibinfo {year} {2008})}\BibitemShut {NoStop}%
\bibitem [{\citenamefont {Shelykh}\ \emph {et~al.}(2010)\citenamefont
  {Shelykh}, \citenamefont {Kavokin}, \citenamefont {Rubo}, \citenamefont
  {Liew},\ and\ \citenamefont {Malpuech}}]{polarization_review}%
  \BibitemOpen
  \bibfield  {author} {\bibinfo {author} {\bibfnamefont {I.~A.}\ \bibnamefont
  {Shelykh}}, \bibinfo {author} {\bibfnamefont {A.~V.}\ \bibnamefont
  {Kavokin}}, \bibinfo {author} {\bibfnamefont {Y.}~\bibnamefont {Rubo}},
  \bibinfo {author} {\bibfnamefont {T.~C.~H.}\ \bibnamefont {Liew}}, \ and\
  \bibinfo {author} {\bibfnamefont {G.}~\bibnamefont {Malpuech}},\ }\href
  {\doibase doi:10.1088/0268-1242/25/1/013001} {\bibfield  {journal} {\bibinfo
  {journal} {Semicond. Sci. Technol.}\ }\textbf {\bibinfo {volume} {25}},\
  \bibinfo {pages} {013001} (\bibinfo {year} {2010})}\BibitemShut {NoStop}%
\bibitem [{\citenamefont {Ciuti}\ \emph {et~al.}(1998)\citenamefont {Ciuti},
  \citenamefont {Savona}, \citenamefont {Piermarocchi}, \citenamefont
  {Quattropani},\ and\ \citenamefont {Schwendimann}}]{Ciuti1998}%
  \BibitemOpen
  \bibfield  {author} {\bibinfo {author} {\bibfnamefont {C.}~\bibnamefont
  {Ciuti}}, \bibinfo {author} {\bibfnamefont {V.}~\bibnamefont {Savona}},
  \bibinfo {author} {\bibfnamefont {C.}~\bibnamefont {Piermarocchi}}, \bibinfo
  {author} {\bibfnamefont {A.}~\bibnamefont {Quattropani}}, \ and\ \bibinfo
  {author} {\bibfnamefont {P.}~\bibnamefont {Schwendimann}},\ }\href {\doibase
  10.1103/PhysRevB.58.7926} {\bibfield  {journal} {\bibinfo  {journal} {Phys.
  Rev. B}\ }\textbf {\bibinfo {volume} {58}},\ \bibinfo {pages} {7926}
  (\bibinfo {year} {1998})}\BibitemShut {NoStop}%
\bibitem [{\citenamefont {Vladimirova}\ \emph {et~al.}(2010)\citenamefont
  {Vladimirova}, \citenamefont {Cronenberger}, \citenamefont {Scalbert},
  \citenamefont {Kavokin}, \citenamefont {Miard}, \citenamefont {Lema\^itre},
  \citenamefont {Bloch}, \citenamefont {Solnyshkov}, \citenamefont {Malpuech},\
  and\ \citenamefont {Kavokin}}]{vlad_interactions}%
  \BibitemOpen
  \bibfield  {author} {\bibinfo {author} {\bibfnamefont {M.}~\bibnamefont
  {Vladimirova}}, \bibinfo {author} {\bibfnamefont {S.}~\bibnamefont
  {Cronenberger}}, \bibinfo {author} {\bibfnamefont {D.}~\bibnamefont
  {Scalbert}}, \bibinfo {author} {\bibfnamefont {K.~V.}\ \bibnamefont
  {Kavokin}}, \bibinfo {author} {\bibfnamefont {A.}~\bibnamefont {Miard}},
  \bibinfo {author} {\bibfnamefont {A.}~\bibnamefont {Lema\^itre}}, \bibinfo
  {author} {\bibfnamefont {J.}~\bibnamefont {Bloch}}, \bibinfo {author}
  {\bibfnamefont {D.}~\bibnamefont {Solnyshkov}}, \bibinfo {author}
  {\bibfnamefont {G.}~\bibnamefont {Malpuech}}, \ and\ \bibinfo {author}
  {\bibfnamefont {A.~V.}\ \bibnamefont {Kavokin}},\ }\href {\doibase
  10.1103/PhysRevB.82.075301} {\bibfield  {journal} {\bibinfo  {journal} {Phys.
  Rev. B}\ }\textbf {\bibinfo {volume} {82}},\ \bibinfo {pages} {075301}
  (\bibinfo {year} {2010})}\BibitemShut {NoStop}%
\bibitem [{\citenamefont {Pethick}\ and\ \citenamefont
  {Smith}(2002)}]{pethick2002}%
  \BibitemOpen
  \bibfield  {author} {\bibinfo {author} {\bibfnamefont {C.}~\bibnamefont
  {Pethick}}\ and\ \bibinfo {author} {\bibfnamefont {H.}~\bibnamefont
  {Smith}},\ }\href@noop {} {\emph {\bibinfo {title} {Bose-Einstein
  condensation in dilute gases}}}\ (\bibinfo  {publisher} {Cambridge University
  Press},\ \bibinfo {year} {2002})\BibitemShut {NoStop}%
\bibitem [{\citenamefont {Solnyshkov}\ \emph {et~al.}(2012)\citenamefont
  {Solnyshkov}, \citenamefont {Flayac},\ and\ \citenamefont
  {Malpuech}}]{soln_monopoles}%
  \BibitemOpen
  \bibfield  {author} {\bibinfo {author} {\bibfnamefont {D.}~\bibnamefont
  {Solnyshkov}}, \bibinfo {author} {\bibfnamefont {H.}~\bibnamefont {Flayac}},
  \ and\ \bibinfo {author} {\bibfnamefont {G.}~\bibnamefont {Malpuech}},\
  }\href {\doibase 10.1103/PhysRevB.85.073105} {\bibfield  {journal} {\bibinfo
  {journal} {Phys. Rev. B}\ }\textbf {\bibinfo {volume} {85}},\ \bibinfo
  {pages} {073105} (\bibinfo {year} {2012})}\BibitemShut {NoStop}%
\bibitem [{\citenamefont {Fischer}\ \emph {et~al.}(2014)\citenamefont
  {Fischer}, \citenamefont {Brodbeck}, \citenamefont {Chernenko}, \citenamefont
  {Lederer}, \citenamefont {Rahimi-Iman}, \citenamefont {Amthor}, \citenamefont
  {Kulakovskii}, \citenamefont {Worschech}, \citenamefont {Kamp}, \citenamefont
  {Durnev}, \citenamefont {Schneider}, \citenamefont {Kavokin},\ and\
  \citenamefont {H\"ofling}}]{spin_meissner}%
  \BibitemOpen
  \bibfield  {author} {\bibinfo {author} {\bibfnamefont {J.}~\bibnamefont
  {Fischer}}, \bibinfo {author} {\bibfnamefont {S.}~\bibnamefont {Brodbeck}},
  \bibinfo {author} {\bibfnamefont {A.~V.}\ \bibnamefont {Chernenko}}, \bibinfo
  {author} {\bibfnamefont {I.}~\bibnamefont {Lederer}}, \bibinfo {author}
  {\bibfnamefont {A.}~\bibnamefont {Rahimi-Iman}}, \bibinfo {author}
  {\bibfnamefont {M.}~\bibnamefont {Amthor}}, \bibinfo {author} {\bibfnamefont
  {V.~D.}\ \bibnamefont {Kulakovskii}}, \bibinfo {author} {\bibfnamefont
  {L.}~\bibnamefont {Worschech}}, \bibinfo {author} {\bibfnamefont
  {M.}~\bibnamefont {Kamp}}, \bibinfo {author} {\bibfnamefont {M.}~\bibnamefont
  {Durnev}}, \bibinfo {author} {\bibfnamefont {C.}~\bibnamefont {Schneider}},
  \bibinfo {author} {\bibfnamefont {A.~V.}\ \bibnamefont {Kavokin}}, \ and\
  \bibinfo {author} {\bibfnamefont {S.}~\bibnamefont {H\"ofling}},\ }\href
  {\doibase 10.1103/PhysRevLett.112.093902} {\bibfield  {journal} {\bibinfo
  {journal} {Phys. Rev. Lett.}\ }\textbf {\bibinfo {volume} {112}},\ \bibinfo
  {pages} {093902} (\bibinfo {year} {2014})}\BibitemShut {NoStop}%
\bibitem [{\citenamefont {Para\"iso}\ \emph {et~al.}(2010)\citenamefont
  {Para\"iso}, \citenamefont {Wouters}, \citenamefont {L\'eger}, \citenamefont
  {Morier-Genoud},\ and\ \citenamefont
  {Deveaud-Pl\'edran}}]{wouters_spin_nature}%
  \BibitemOpen
  \bibfield  {author} {\bibinfo {author} {\bibfnamefont {T.~K.}\ \bibnamefont
  {Para\"iso}}, \bibinfo {author} {\bibfnamefont {M.}~\bibnamefont {Wouters}},
  \bibinfo {author} {\bibfnamefont {Y.}~\bibnamefont {L\'eger}}, \bibinfo
  {author} {\bibfnamefont {F.}~\bibnamefont {Morier-Genoud}}, \ and\ \bibinfo
  {author} {\bibfnamefont {B.}~\bibnamefont {Deveaud-Pl\'edran}},\ }\href
  {http://dx.doi.org/10.1038/nmat2787} {\bibfield  {journal} {\bibinfo
  {journal} {Nat Mater}\ }\textbf {\bibinfo {volume} {9}},\ \bibinfo {pages}
  {655} (\bibinfo {year} {2010})}\BibitemShut {NoStop}%
\bibitem [{\citenamefont {Wouters}\ and\ \citenamefont
  {Carusotto}(2007{\natexlab{a}})}]{michiel_probe}%
  \BibitemOpen
  \bibfield  {author} {\bibinfo {author} {\bibfnamefont {M.}~\bibnamefont
  {Wouters}}\ and\ \bibinfo {author} {\bibfnamefont {I.}~\bibnamefont
  {Carusotto}},\ }\href {\doibase 10.1103/PhysRevA.76.043807} {\bibfield
  {journal} {\bibinfo  {journal} {Phys. Rev. A}\ }\textbf {\bibinfo {volume}
  {76}},\ \bibinfo {pages} {043807} (\bibinfo {year}
  {2007}{\natexlab{a}})}\BibitemShut {NoStop}%
\bibitem [{\citenamefont {Berceanu}\ \emph {et~al.}(2012)\citenamefont
  {Berceanu}, \citenamefont {Cancellieri},\ and\ \citenamefont
  {Marchetti}}]{andrei_drag}%
  \BibitemOpen
  \bibfield  {author} {\bibinfo {author} {\bibfnamefont {A.~C.}\ \bibnamefont
  {Berceanu}}, \bibinfo {author} {\bibfnamefont {E.}~\bibnamefont
  {Cancellieri}}, \ and\ \bibinfo {author} {\bibfnamefont {F.~M.}\ \bibnamefont
  {Marchetti}},\ }\href {http://stacks.iop.org/0953-8984/24/i=23/a=235802}
  {\bibfield  {journal} {\bibinfo  {journal} {Journal of Physics: Condensed
  Matter}\ }\textbf {\bibinfo {volume} {24}},\ \bibinfo {pages} {235802}
  (\bibinfo {year} {2012})}\BibitemShut {NoStop}%
\bibitem [{\citenamefont {Landau}\ and\ \citenamefont
  {Lifshitz}(1987)}]{landau_lifshitz_FD}%
  \BibitemOpen
  \bibfield  {author} {\bibinfo {author} {\bibfnamefont {L.~D.}\ \bibnamefont
  {Landau}}\ and\ \bibinfo {author} {\bibfnamefont {E.~M.}\ \bibnamefont
  {Lifshitz}},\ }\href {http://www.worldcat.org/isbn/0750627670} {\emph
  {\bibinfo {title} {Fluid Mechanics, Second Edition: Volume 6 (Course of
  Theoretical Physics)}}},\ \bibinfo {edition} {2nd}\ ed.,\ Course of
  theoretical physics / by L. D. Landau and E. M. Lifshitz, Vol. 6\ (\bibinfo
  {publisher} {Butterworth-Heinemann},\ \bibinfo {year} {1987})\BibitemShut
  {NoStop}%
\bibitem [{\citenamefont {Wouters}\ and\ \citenamefont
  {Carusotto}(2007{\natexlab{b}})}]{michiel_OPO}%
  \BibitemOpen
  \bibfield  {author} {\bibinfo {author} {\bibfnamefont {M.}~\bibnamefont
  {Wouters}}\ and\ \bibinfo {author} {\bibfnamefont {I.}~\bibnamefont
  {Carusotto}},\ }\href {\doibase 10.1103/PhysRevB.75.075332} {\bibfield
  {journal} {\bibinfo  {journal} {Phys. Rev. B}\ }\textbf {\bibinfo {volume}
  {75}},\ \bibinfo {pages} {075332} (\bibinfo {year}
  {2007}{\natexlab{b}})}\BibitemShut {NoStop}%
\bibitem [{\citenamefont {Solnyshkov}\ \emph {et~al.}(2009)\citenamefont
  {Solnyshkov}, \citenamefont {Shelykh},\ and\ \citenamefont
  {Malpuech}}]{phase_disorder}%
  \BibitemOpen
  \bibfield  {author} {\bibinfo {author} {\bibfnamefont {D.}~\bibnamefont
  {Solnyshkov}}, \bibinfo {author} {\bibfnamefont {I.}~\bibnamefont {Shelykh}},
  \ and\ \bibinfo {author} {\bibfnamefont {G.}~\bibnamefont {Malpuech}},\
  }\href {\doibase 10.1103/PhysRevB.80.165329} {\bibfield  {journal} {\bibinfo
  {journal} {Phys. Rev. B}\ }\textbf {\bibinfo {volume} {80}},\ \bibinfo
  {pages} {165329} (\bibinfo {year} {2009})}\BibitemShut {NoStop}%
\bibitem [{\citenamefont {Rubo}\ \emph {et~al.}(2006)\citenamefont {Rubo},
  \citenamefont {Kavokin},\ and\ \citenamefont
  {Shelykh}}]{suppression_superfluidity}%
  \BibitemOpen
  \bibfield  {author} {\bibinfo {author} {\bibfnamefont {Y.~G.}\ \bibnamefont
  {Rubo}}, \bibinfo {author} {\bibfnamefont {A.}~\bibnamefont {Kavokin}}, \
  and\ \bibinfo {author} {\bibfnamefont {I.}~\bibnamefont {Shelykh}},\ }\href
  {\doibase http://dx.doi.org/10.1016/j.physleta.2006.05.015} {\bibfield
  {journal} {\bibinfo  {journal} {Physics Letters A}\ }\textbf {\bibinfo
  {volume} {358}},\ \bibinfo {pages} {227 } (\bibinfo {year}
  {2006})}\BibitemShut {NoStop}%
\end{thebibliography}

%

\end{document}